\documentclass[journal,twocolumn,10pt]{IEEEtran} % double column submit

\usepackage{cite}
\ifCLASSINFOpdf
  \usepackage[pdftex]{graphicx}
  \usepackage{epstopdf} % remove this
\else
\fi
\usepackage{amsmath}
\usepackage{algorithm}
\usepackage{algorithmic}

  \usepackage[caption=false,font=footnotesize]{subfig}
\usepackage{url}
\usepackage{array}

\usepackage{amsthm}

\newtheoremstyle{myremark}% name of the style to be used
{}% measure of space to leave above the theorem. E.g.: 3pt
{}% measure of space to leave below the theorem. E.g.: 3pt
{}% name of font to use in the body of the theorem
{0pt}% measure of space to indent
{\bfseries}% name of head font
{.}% punctuation between head and body
{ }% space after theorem head; " " = normal interword space
{\thmname{#1}\thmnumber{ #2}: \thmnote{#3}}

\theoremstyle{theoremdd}

\theoremstyle{myremark}

\newtheoremstyle{myshortremark}% name of the style to be used
{}% measure of space to leave above the theorem. E.g.: 3pt
{}% measure of space to leave below the theorem. E.g.: 3pt
{}% name of font to use in the body of the theorem
{0pt}% measure of space to indent
{\bfseries}% name of head font
{.}% punctuation between head and body
{ }% space after theorem head; " " = normal interword space
{\thmname{#1}\thmnumber{ #2}: \thmnote{#3}}

\theoremstyle{myshortremark}
\newcounter{take}
\newtheorem{takeaway}[take]{Takeaway}

\newtheoremstyle{mydefinition}% name of the style to be used
{}% measure of space to leave above the theorem. E.g.: 3pt
{}% measure of space to leave below the theorem. E.g.: 3pt
{}% name of font to use in the body of the theorem
{0pt}% measure of space to indent
{\bfseries}% name of head font
{.}% punctuation between head and body
{ }% space after theorem head; " " = normal interword space
{\thmname{#1}\thmnumber{ #2}: \thmnote{#3}}
\theoremstyle{mydefinition}
\newtheorem{definition}{Definition}

\newcommand{\comment}[1]{{}} % used to comment stuff out easily

\usepackage{amssymb}
\usepackage{xspace}
\usepackage{bm}
\usepackage{bbm}

\def\b0{{\mathbf{0}}}

\renewcommand{\vec}[1]{{\mathbf{#1}}\xspace} % vectors

\newcommand{\parens}[1]{{\left(#1\right)}\xspace}

\newcommand{\braces}[1]{{\left\{#1\right\}}\xspace}
\newcommand{\bars}[1]{{\left\vert#1\right\vert}\xspace}

\newcommand{\inv}{\ensuremath{^{-1}}\xspace}

\newcommand{\logtwo}[1]{\ensuremath{\mathrm{log}_{2}\parens{#1}}}
\newcommand{\logten}[1]{\ensuremath{\mathrm{log}_{10}\parens{#1}}}

\newcommand{\card}[1]{\bars{#1}}

\newcommand{\st}{\ensuremath{\mathrm{s.t.~}}\xspace}
\newcommand{\opt}{\ensuremath{^{\star}}\xspace}

\newcommand{\msnr}{\ensuremath{\mathsf{SNR}}\xspace}

\newcommand{\msinr}{\ensuremath{\mathsf{SINR}}\xspace}

\newcommand{\minr}{\ensuremath{\mathsf{INR}}\xspace}

\newcommand{\minrth}{\ensuremath{\mathsf{INR}_\mathrm{th}}\xspace}

\newcommand{\powertx}{{P_{\mathrm{tx}}}}
\newcommand{\powernoise}{{P_{\mathrm{n}}}}

\newcommand{\powerint}{\ensuremath{P_{\mathrm{int}}}\xspace}

\newcommand{\Gtx}{{G_{\mathrm{tx}}}}
\newcommand{\Grx}{{G_{\mathrm{rx}}}}

\newcommand{\setsatp}{\mathcal{P}}
\newcommand{\setsats}{\mathcal{S}}

\newcommand{\msnrup}{\msnr(\sfu; \vp)}

\newcommand{\minrups}{\minr(\sfu, \vp; \vs)}

\newcommand{\minrvps}{\minr(\sfv, \vp; \vs)}

\newcommand{\minrupopts}{\minr(\sfu, \vp\opt; \vs)}

\newcommand{\minrmaxu}{\minr_{\mathrm{max}}(\sfu)}

\newcommand{\minrminu}{\minr_{\mathrm{min}}(\sfu)}

\newcommand{\minrmaxupopt}{\minr_{\mathrm{max}}(\sfu, \vp\opt)}

\newcommand{\minrminupopt}{\minr_{\mathrm{min}}(\sfu, \vp\opt)}

\def\sfu{{\mathtt{u}}}
\def\sfv{{\mathtt{v}}}

\def\vp{{\vec{p}}}

\def\vs{{\vec{s}}}

\usepackage{xcolor}
\newcommand{\red}[1]{\textcolor{black}{#1}}

\newcommand{\edit}[1]{\textcolor{black}{#1}}

\usepackage{xspace}
\usepackage[acronym,nogroupskip,nonumberlist,nopostdot]{glossaries}
\makeglossaries

\newacronym{snr}{SNR}{signal-to-noise ratio}
\newacronym{sinr}{SINR}{signal-to-interference-plus-noise ratio}
\newacronym{sir}{SIR}{signal-to-interference ratio}
\newacronym{inr}{INR}{interference-to-noise ratio}
\newacronym{pdf}{PDF}{probability distribution function}
\newacronym{cdf}{CDF}{cumulative distribution function}
\newacronym{leo}{LEO}{low-earth orbit}
\newacronym{ngso}{NGSO}{non-geostationary orbit}
\newacronym{pu}{PU}{primary user}
\newacronym{su}{SU}{secondary user}
\newacronym{frf}{FRF}{frequency reuse factor}
\newacronym{los}{LOS}{line-of-sight}
\newacronym{nlos}{NLOS}{non-line-of-sight}
\newacronym{mimo}{MIMO}{multiple-input multiple-output}
\newacronym{sr}{SR}{Shadowed Rician}
\newacronym{ssr}{SSR}{Squared Shadowed Rician}
\newacronym{5g}{5G}{fifth generation}
\newacronym{eirp}{EIRP}{effective isotropic radiated power}
\newacronym{fcc}{FCC}{Federal Communications Commission}
\newacronym{itu}{ITU}{International Telecommunication Union}
\newcommand{\snr}{\gls{snr}\xspace}

\newcommand{\inr}{\gls{inr}\xspace}

\newcommand{\leo}{\gls{leo}\xspace}

\newcommand{\gcdf}{\gls{cdf}\xspace}
\newcommand{\gpcdf}{\glspl{cdf}\xspace}

\newcommand{\gsnr}{\gls{snr}\xspace}
\newcommand{\ginr}{\gls{inr}\xspace}
\newcommand{\gsinr}{\gls{sinr}\xspace}

\newcommand{\gpsnr}{\glspl{snr}\xspace}

\newcommand{\gpsinr}{\glspl{sinr}\xspace}

\newcommand{\tabref}[1]{Table~\ref{#1}}
\newcommand{\figref}[1]{\figurename~\ref{#1}}

\usepackage{mathtools}

\begin{document}

%
% paper title
% Titles are generally capitalized except for words such as a, an, and, as,
% at, but, by, for, in, nor, of, on, or, the, to and up, which are usually
% not capitalized unless they are the first or last word of the title.
% Linebreaks \\ can be used within to get better formatting as desired.
% Do not put math or special symbols in the title.
% \title{Feasibility Studies on the Coexistence of Heterogeneous LEO Satellite Communication Systems}
% \title{On the Feasibility of In-Band Coexistence\\between Heterogeneous Low-Earth Orbit\\Satellite Communication Systems}
\title{Feasibility Analysis of In-Band Coexistence in Dense LEO Satellite Communication Systems}
%
%
% author names and IEEE memberships
% note positions of commas and nonbreaking spaces ( ~ ) LaTeX will not break
% a structure at a ~ so this keeps an author's name from being broken across
% two lines.
% use \thanks{} to gain access to the first footnote area
% a separate \thanks must be used for each paragraph as LaTeX2e's \thanks
% was not built to handle multiple paragraphs
%

\author{%
    Eunsun Kim,~%
	Ian~P.~Roberts,~%
	and Jeffrey~G.~Andrews%
    \thanks{E.~Kim and J.~G.~Andrews are with the 6G@UT Research Center and the Wireless Networking and Communications Group at the University of Texas at Austin. I.~P.~Roberts is with the Wireless Lab at UCLA.}
    % \thanks{Updated: \today.}
	% \thanks{E.~Kim and J.~G.~Andrews are with the 6G@UT Research Center and the Wireless Networking and Communications Group at the University of Texas at Austin, Austin, TX 78712 USA. I.~P.~Roberts is with the Wireless Lab at UCLA. Updated: \today.}
}

\maketitle

\begin{abstract}
This work provides a rigorous methodology for assessing the feasibility of spectrum sharing between large \leo satellite constellations.  For concreteness, we focus on the existing Starlink system and the soon-to-be-launched Kuiper system, which is prohibited from inflicting excessive interference onto the incumbent Starlink ground users.
We carefully model and study the potential downlink interference between the two systems at 20~GHz and investigate how strategic satellite selection may be used by Kuiper to serve its own ground users while also protecting Starlink ground users. We then extend this notion of satellite selection to the case where Kuiper has limited knowledge of Starlink's serving satellite. Throughout our analysis, we examine the distribution of interference and SINR as each constellation orbits the globe.
Our findings reveal that there is nearly always the potential for very high and extremely low interference, depending on which Starlink and Kuiper satellites are being used to serve their ground users.
Consequently, we show that Kuiper can protect Starlink ground users with high probability, by strategically selecting which of its satellites are used to serve its ground users. 
Simultaneously, Kuiper is capable of delivering near-maximal downlink SINR to its own ground users.
This highlights a feasible route to the coexistence of two dense \leo satellite systems, even in scenarios where one system has limited knowledge of the other's serving satellites.  
\end{abstract}
\vspace{-0.12cm}

%This work provides a rigorous methodology for assessing the feasibility of spectrum sharing between large low-earth orbit (LEO) satellite constellations. For concreteness, we focus on the existing Starlink system and the soon-to-be-launched Kuiper system, which is prohibited from inflicting excessive interference onto the incumbent Starlink ground users. We carefully model and study the potential downlink interference between the two systems and investigate how strategic satellite selection may be used by Kuiper to serve its ground users while also protecting Starlink ground users. We then extend this notion of satellite selection to the case where Kuiper has limited knowledge of Starlink's serving satellite. Our findings reveal that there is always the potential for very high and extremely low interference, depending on which Starlink and Kuiper satellites are being used to serve their users. Consequently, we show that Kuiper can protect Starlink ground users with high probability, by strategically selecting which of its satellites are used to serve its ground users. Simultaneously, Kuiper is capable of delivering near-maximal downlink SINR to its own ground users. This highlights a feasible route to the coexistence of two dense LEO satellite systems, even in scenarios where one system has limited knowledge of the other's serving satellites.  

%\input{sec-abstract-JGA.tex}

%\input{sec-peer-review-title.tex}

\glsresetall % resets glossary terms after abstract, comment out if desired
%\newpageth
%\tableofcontents

%\input{literature_review.tex}

%\input{sec-introduction-v02.tex}

\section{Introduction}
A new paradigm of global broadband connectivity is unfolding as mega-constellations comprised of thousands of \leo satellites %aim
are deployed to deliver wireless coverage across the globe.
These \leo satellite communication systems circumvent the time-consuming challenges associated with deploying terrestrial infrastructure, which has left many regions and communities % being 
severely under-served---or even completely unserved---by terrestrial-based modes of connectivity \cite{toward_6g}. %,3gpp22822}. 
Two of the most notable large-scale efforts in this pursuit are SpaceX's Starlink \cite{McDowell_2020,spaceX_req}, with over 5,000 satellites currently in orbit, and Amazon's Project Kuiper \cite{kuiper}, %, kuiper_koller}, 
which recently launched test satellites in October 2023 \cite{sat_stats} and intends to launch 3,236 satellites throughout the decade.

These mega-constellations have already transformed the connectivity landscape and will be a key new source of broadband access in the 6G era \cite{sat_integ, sat_6g, path_6g}.
However, among other concerns, there are important open questions regarding the coexistence of multiple mega-constellations \cite{cr_sat1, leo_6g, kasat_cr, dense_leo}. %satcom_challenges,kasat_cr, dense_leo}. 
Unlike in traditional terrestrial cellular networks, frequency spectrum has been allocated to these satellite systems in a \textit{non-exclusive} manner by the \gls{fcc} in the United States \edit{and other global spectrum regulators \cite{47cfr, itur1323}.} 
To facilitate fair spectrum sharing, the \gls{fcc} gives precedence (or incumbency) to systems which applied for launch rights in earlier so-called processing rounds than others.
Consequently, it is the expectation of the \gls{fcc} that each system either coordinate with or protect systems which acquired launch rights in earlier processing rounds.

Given the sheer number of satellites slated to launch and the growing number requesting launch rights  \cite{osstp}, spectrum sharing across these satellite systems is far from trivial, especially when one considers the dynamics over time as satellites orbit.
The feasibility of this coexistence is currently unclear and mechanisms to facilitate \edit{coexistence} % such 
are even less established.
In this work, we take a necessary first step by carefully investigating the severity of in-band interference over time between coexisting \leo satellite systems as their constellations orbit the globe, which ultimately sheds light on the efficacy of potential mechanisms to protect incumbent systems---and how imposing this protection may impact system performance.
\vspace{-0.12cm}

\subsection{Related Work and Regulations}

For decades, spectrum sharing has been investigated from both academic and regulatory perspectives with various applications in mind \cite{fcc_cr,satcom_cr}.
%In the context of satellite systems, 
The principle of spectrum sharing is concerned with so-called \textit{secondary} systems not inflicting significant interference onto \textit{primary} (or incumbent) systems when attempting to access some portion of frequency spectrum.
Cognitive radio in the context of satellite systems has been proposed to manage and mitigate this interference via mechanisms including spectrum sensing, underlay, overlay, and database methods \cite{satcom_cr_overlay, satcom_cr1, cr_sat1, database}.
The basic idea of \textit{underlay} methods is that, when primary systems are deemed idle, a secondary system can use the free spectrum opportunistically \cite{why_cr_satcom, ss, css}. 
On the other hand, \textit{overlay} techniques allow a secondary system to use the spectrum concurrently with primary systems, assuming the secondary system does not substantially impact normal operation of the primary systems---which naturally leads to discussions and debate on {what defines an acceptable level of interference}. 
Various mechanisms along multiple dimensions have been studied to enable such overlay coexistence, with power allocation \cite{why_cr_satcom, pc_1, pc_2} and spatial domain beamforming \cite{beam1,  beam3} being two prominent proposed routes to protect primary systems.

Regulatory bodies play a key role in establishing clear and comprehensive rules for spectrum sharing to realize coexistence between \leo satellite systems.
The \gls{fcc}, for instance, has employed an overlay coexistence paradigm, allowing secondary systems to inflict marginal interference onto primary systems, since this can facilitate more efficient and more widespread use of spectral resources \cite{why_cr_satcom, css_mb}.
Defining the level at which interference becomes prohibitive to a primary system has proven to be a complicated task involving a variety of priorities from multiple perspectives, and as a result, it has been difficult to formulate and regulate a so-called \textit{protection constraint} which the secondary system must oblige \cite{ntia_ipc}.

This protection constraint in satellite systems has been often formulated as a threshold on the interference power relative to the noise floor of the primary system, i.e., the \gls{inr}.
For instance, it has been proposed that the resulting \ginr not be more than a threshold ranging from $-6$~dB to an even stricter $-12.2$~dB, perhaps for some specified fraction of time \cite{ntia_ipc, itur1155, itur1432,itur1323}.
Beyond looking purely at interference, a constraint on throughput degradation has also been considered to more meaningfully quantify the impact of interference on the primary system \cite{fnprm}.
This is naturally more difficult to quantify from the perspective of the secondary system, since it depends on the signal quality of the primary system, and thus introduces open questions on how to satisfy and evaluate such a protection constraint.
In this work, we will consider a pure \ginr constraint for concreteness.
Also, it should be noted that a throughput degradation constraint can be suitably mapped to an interference power constraint for a given primary system signal quality.

\vspace{-0.12cm}
\subsection{Contributions}

In this paper, we analyze the coexistence paradigm laid forth thus far: the operation of a \textit{primary} (or incumbent) \leo satellite system in the presence of a \textit{secondary} \leo satellite system attempting to coexist in-band under the expectation that it protects the primary system.
In light of active discussions in the regulatory domain and the impending launch of thousands of additional \leo satellites, this paper aims to characterize downlink interference and investigate the feasibility of the secondary system to reliably protect the primary system. We accomplish this by creating a high-fidelity simulation of the two \leo satellite systems using actual orbital parameters, transmit powers, antenna gains, and other system parameters reported in public \gls{fcc} filings \cite{spaceX_ss, kuiper_ss}. Our aim is to shed light on the feasibility of coexistence in \leo satellite communication systems and to motivate future work proposing mechanisms to enable such coexistence. 
To accomplish this, we broadly investigate three important questions surrounding this coexistence between satellite systems.

   \textbf{Section IV: How much interference will be inflicted by the secondary system onto the primary?} 
    Perhaps the most fundamental concern we explore in this work is quantifying how much interference a secondary satellite may inflict onto a primary ground user. 
    To do this, we evaluate the absolute bounds on this interference, as well as the bounds when conditioning on a particular primary satellite serving the ground user.
    We examine the distributions of both of these bounds over time as the satellites orbit around the globe and conclude that interference can be very high or extremely low at virtually any given time, depending heavily on the pair of primary and secondary satellites being used to serve downlink.
    
   % \item 
   \textbf{Section V: Can the secondary system protect the primary and what does it sacrifice in doing so?}
    We then investigate how feasible it is for the secondary system to reliably protect a primary user under various interference constraints.
    In addition to mere feasibility, we also examine the number of secondary satellites capable of meeting this constraint.
    We then illustrate that strategic satellite selection may be used by the secondary system to serve its own ground user while also protecting a primary ground user.
    In doing so, we dissect the sacrifices in secondary system performance under such a technique, in terms of received \gsinr.
    Again, we show all of this in distribution across the constellations' orbits and remarkably discover that the secondary system often must only sacrifice a fraction of a decibel in \gsinr when protecting primary ground users.

   \textbf{Section VI: How much knowledge does the secondary system need about the primary?}
    Finally, we conclude by reexamining the viability of protection when the secondary system has limited knowledge about which primary serving satellite is serving a given ground user.
    We observe that, on average, there are 10 secondary satellites capable of guaranteeing that primary ground users are protected, even when the secondary system has only modest knowledge about the vicinity of the primary serving satellite.
    While more uncertainty and a more stringent protection constraint certainly makes it more difficult to guarantee protection, we impressively find that the secondary system only sacrifices at most around 2 dB of \gsinr on average and around 4 dB in the worst case.
    
\vspace{-0.12cm}

\begin{figure}[t]
    \centering
    \includegraphics[width=\linewidth,height=0.255\textheight,keepaspectratio]{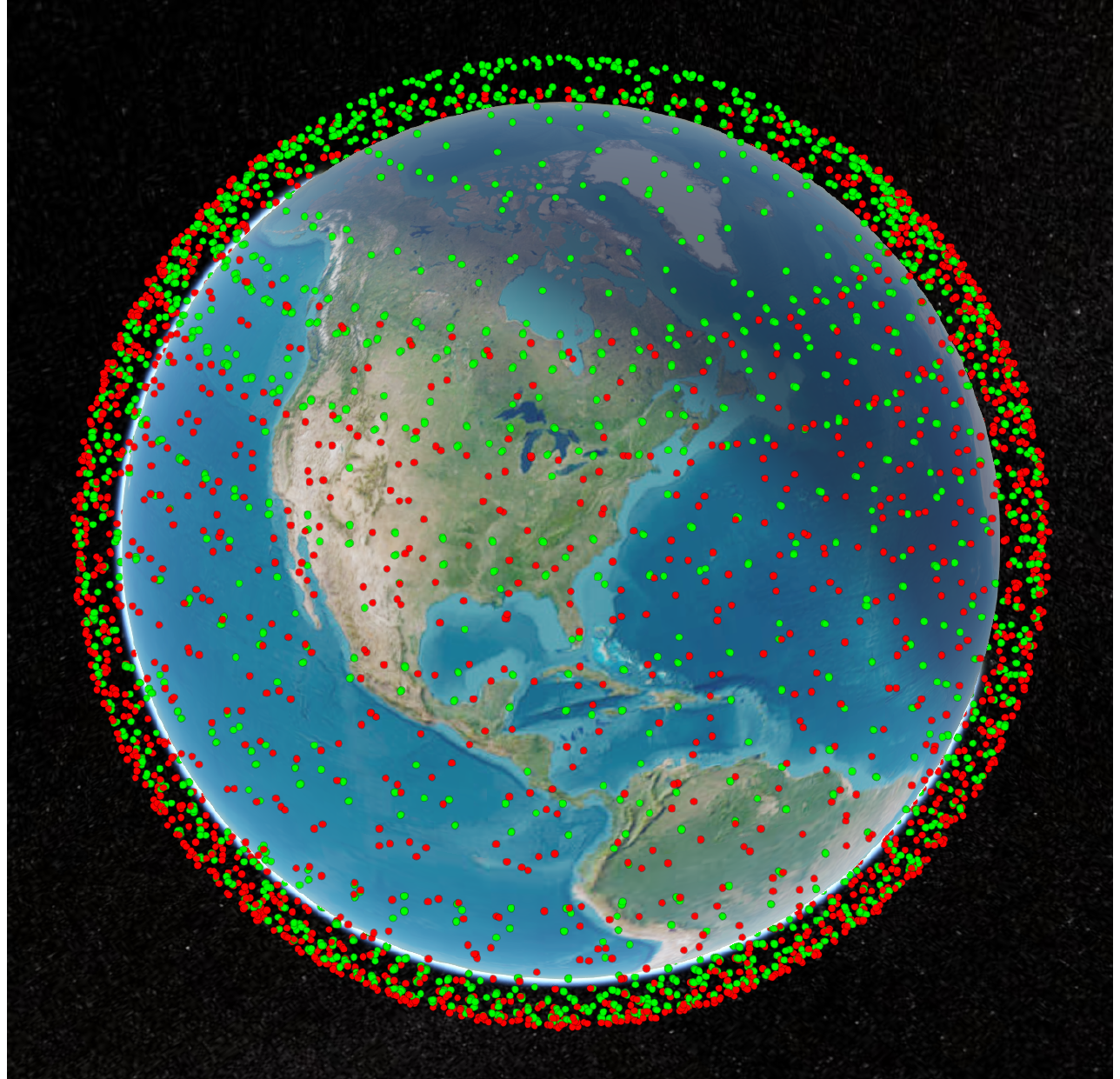}
    \caption{The satellites in SpaceX's Starlink constellation (green) and Amazon's Project Kuiper constellation (red) based on public filings \cite{spaceX_req, spaceX_grt, kuiper}. This paper investigates how these systems may interfere with one another when operating in the same frequency band and the feasibility of their coexistence.}
    \label{fig:constellation}
    \vspace{-0.25cm}
 \end{figure}

\section{System Model} \label{sec:system-model}

This work considers two downlink \leo satellite communication systems, each operating independently to serve % downlink to a corresponding 
ground users on the surface of the Earth.
Each of the two systems is comprised of a unique {constellation} of satellites orbiting 
at altitudes on the order of 500~km. 
We term one of the systems the \textit{primary} system and the other the \textit{secondary} system.
The work herein is interested in assessing the in-band interference inflicted onto a given ground user of one system by a single satellite of the other system.
As such, we consider the case where the two systems are transmitting downlink to nearby ground users at the same time and at the same carrier frequency of 20 GHz. In both systems, we assume each satellite employs a high-gain antenna (e.g., phased array or dish) to form a highly directional beam in the general direction of a ground user it aims to serve.

\begin{figure}[!t]
	\centering
	\includegraphics[width=\linewidth,height=0.255\textheight,keepaspectratio]{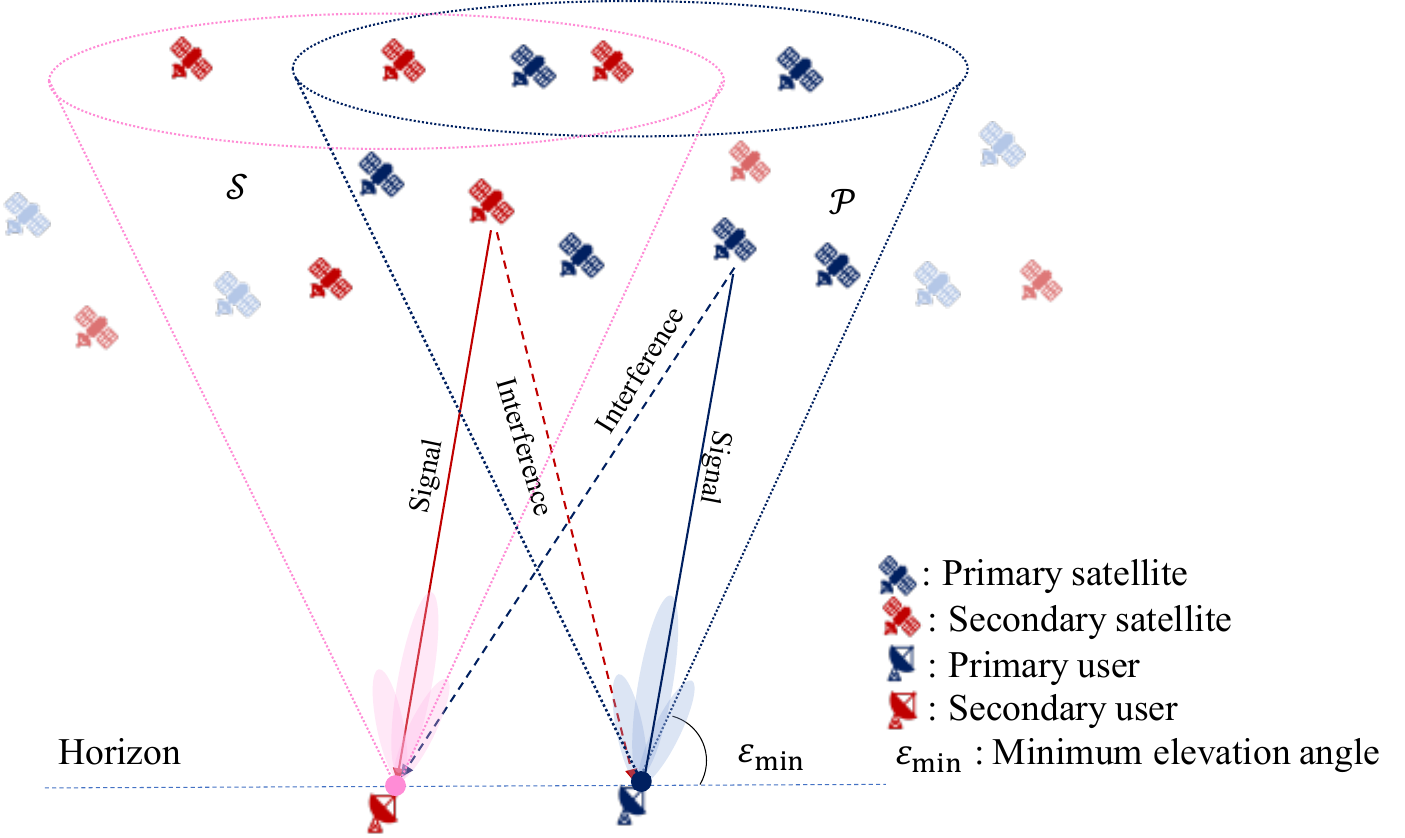}
	\caption{This work considers a scenario where a primary satellite system and a secondary satellite system interfere with one another when attempting to serve downlink to ground users at the same time and at the same frequency. The interference inflicted onto a primary ground user depends on its own receive beam and on the transmit beam of an interfering secondary satellite. }
	\label{fig:system}
%	\vspace{-0.25cm}
\end{figure}

Let us define the transmit antenna gain of the primary satellite $\vp$ in the direction of a ground user $\sfu$ as $\Gtx(\sfu,\vp)$.
Similarly, let $\Grx(\sfu,\vp)$ be the receive antenna gain of user $\sfu$ in the direction of primary satellite $\vp$.
The path loss between the two we denote as $L(\sfu,\vp)$.
With this, we can define the received \gsnr at a primary ground user $\sfu$ served by its serving satellite $\vp$ as
\begin{align}
\msnr(\sfu,\vp) = \frac{\powertx(\vp) \ \Gtx(\sfu,\vp) \ \Grx(\sfu,\vp) \ L(\sfu,\vp)\inv}{\powernoise(\sfu)},
\end{align}
where $\powertx(\vp)$ is the transmit power of $\vp$ and $\powernoise(\sfu)$ is the noise power at $\sfu$.
Generalizing this notation to the secondary system, the received \snr at a secondary system ground user $\sfv$ from its satellite $\vs$ is
\begin{align}
\msnr(\sfv,\vs) = \frac{\powertx(\vs) \ \Gtx(\sfv,\vs) \ \Grx(\sfv,\vs) \ L(\sfv,\vs)\inv}{\powernoise(\sfv)}.
\end{align}

As mentioned, we are particularly interested in the interference inflicted onto each system by the other. 
In this work, our focus is exclusively on the interference inflicted by individual satellites from each system, in order to draw concrete conclusions on the interplay between two satellites. 
Other sources of interference, such as from other beams \cite{our_tvt_sr} or other satellites across the constellations, is certainly a worthwhile direction for future work.%
\footnote{Multi-beam satellites serving multiple ground users at once are not considered explicitly in this work to focus on the impact of interference from a single pair of beams from a single pair of satellites.} 
Suppose a secondary satellite $\vs$ inflicts interference onto a primary ground user $\sfu$ being served by a primary satellite $\vp$.
% The resulting \inr captures the strength of this interference relative to the noise power and takes the following form.
% The resulting \inr at the primary user $\sfu$ served by satellite $\vp$ in the presence of a secondary satellite $\vs$ is \edit{kind of repetitive}
Then, the resulting \inr at the primary ground user $\sfu$ is
\begin{align} \label{eq:inr-p}
\minr(\sfu,\vp;\vs) = \frac{\powertx(\vs) \ \Gtx(\sfu,\vs;\sfv) \ \Grx(\sfu,\vs;\vp) \ L(\sfu,\vs)\inv}{\powernoise(\sfu)}.
\end{align}
Here, we have slightly extended the notation of $\Gtx(\cdot)$ and $\Grx(\cdot)$ to make it clear that $\Gtx(\sfu,\vs;\sfv)$ represents the transmit gain of the secondary satellite $\vs$ in the direction of the primary user $\sfu$ when $\vs$ serves its secondary ground user $\sfv$.
Similarly, $\Grx(\sfu,\vs;\vp)$ represents the receive gain in the direction of the secondary satellite $\vs$ when the ground user $\sfu$ steers its antenna toward its serving satellite $\vp$.
By virtue of this, the primary ground user's \ginr depends on its primary serving satellite $\vp$ and on the secondary satellite $\vs$.
A primary satellite $\vp$ likewise will inflict interference onto a secondary ground user $\sfv$ being served by satellite $\vs$, leading to an \ginr \edit{at $\sfv$} of
\begin{align}
\minr(\sfv,\vs;\vp) = \frac{\powertx(\vp) \ \Gtx(\sfv,\vp;\sfu) \ \Grx(\sfv,\vp;\vs) \ L(\sfv,\vp)\inv}{\powernoise(\sfv)}.
\end{align}

Together, the received \gsnr and \ginr at the primary ground user $\sfu$ and at the secondary ground user $\sfv$ dictate their \gpsinr, which take the familiar forms
\begin{align}
\msinr(\sfu,\vp;\vs) = \frac{\msnr(\sfu,\vp)}{1 + \minr(\sfu,\vp;\vs)},\\% \qquad
 \msinr(\sfv,\vs;\vp) = \frac{\msnr(\sfv,\vs)}{1 + \minr(\sfv,\vs;\vp)}.
\end{align}
A prohibitively high \ginr (e.g., $\minr \gg 0$ dB) would lead to significant degradation in link quality where $\msinr \ll \msnr$.
As such, we are motivated to investigate the severity of interference in the sequel, but first we lay the groundwork on the methodology by which we accomplish such an investigation.

% ---

% save this for methodology
% We assume this is accomplished by steering toward center of cell.

\section{Methodology of Our Feasibility Analysis} \label{sec:methodology}

To conduct a thorough analysis on the feasibility of two dense \leo satellite communication systems coexisting in-band \edit{at 20 GHz}, we consider the two preeminent commercial systems mentioned in the introduction: Starlink by SpaceX  %\cite{McDowell_2020} 
and Project Kuiper by Amazon. % \cite{kuiper_koller}.
We consider Starlink as the primary system and Kuiper as the secondary system; this is motivated by the fact that Starlink has priority rights to the $19.7$--$20.2$~GHz band for downlink transmission \cite{spaceX_req}.
Indeed, based on current regulations, Kuiper is permitted to operate within that band along with Starlink, with an understanding that it not cause prohibitive interference to Starlink \cite{kuiper}---\edit{coinciding with} %precisely the motivation behind 
the analyses herein.

\subsection{LEO Satellite Constellations}
We simulate the constellation of satellites in Starlink and Kuiper in a Walker-Delta fashion \cite{wd} %, wd1} 
based on the orbital parameters detailed in \tabref{tab:spaceX} and \tabref{tab:kuiper}, which are extracted from public filings \cite{kuiper, spaceX_req, spaceX_grt}.\footnote{This work considers the first-generation Starlink constellation.} 
The two constellations can be seen overlayed one another in \figref{fig:constellation}.
Let $\bar{\mathcal{P}} = \braces{\vp_i : i=1,\dots,4408}$ be the set of all $4408$ satellites in Starlink's constellation.
Similarly, let $\bar{\mathcal{S}} = \braces{\vs_i : i=1,\dots,3236}$ be the set of all $3236$ satellites in Kuiper's constellation.
At a given instant in time, the systems aim to serve some primary ground user $\sfu$ and secondary ground user~$\sfv$. 

\begin{table}[t]
    \caption{SpaceX's Starlink Constellation Parameters \cite{spaceX_req,spaceX_grt}}
    \centering
    \label{tab:spaceX}
    \begin{tabular}{| c| c|c|c|c| }
        \hline
        % Alt. ($km$) & Inc. (${}^\circ$)  & $\mathrm{N_{pl}}$ & $\mathrm{N_{sat-pl}}$ & $\mathrm{N_{sat-total}}$  \\
        Altitude & Inclination &  Planes & Satellites/Plane & Total No.~Satellites \\
        \hline
        $540$ km & $ 53.2^\circ$ &$ 72$ & $22$& $1584$ \\
        \hline
        $550$ km & $ 53^\circ$ & $72$ & $22$ & $1584$ \\
        \hline 
        $560$ km & $97.6^\circ$ & $4$ & $43$ & $172$ \\
        \hline 
        $560$ km & $97.6^\circ$ & $6$ & $58$ & $348$ \\
        \hline
        $570$ km & $70^\circ$ &$ 36$ &$ 20$ &$ 720$ \\
        \hline
    \end{tabular}
%\vspace{-0.25cm}
\end{table}

\begin{table}[t]
    \caption{Amazon's Project Kuiper Constellation Parameters \cite{kuiper}}
    \centering
    \label{tab:kuiper}
    \begin{tabular}{| c| c|c|c|c| }
        \hline
        Altitude & Inclination &  Planes & Satellites/Plane & Total No.~Satellites \\
         \hline
        $590$ km & $33^\circ$ & $28$ & $28$ &$ 784$ \\
        \hline 
        $610$ km & $42^\circ$ & $36$ & $36$ & $1296$ \\
        \hline 
        $630$ km & $51.9^\circ$ & $34$ & $34$& $1156$ \\
        \hline
    \end{tabular}
% \vspace{-0.25cm}
\end{table}

When aiming to serve these users in practice, each of these systems will select one satellite within its set of \textit{overhead} satellites.
More specifically, each system will select a satellite that is within some cone centered on the vector normal to the surface of the Earth at the ground user's location, as shown in \figref{fig:system}. 
In practice, this set of overhead satellites is typically defined by a minimum elevation angle $\epsilon_{\mathrm{min}} > 0$ relative to the user's horizon, at or above which a satellite must lie to be considered for selection.
It is reasonable to assume both systems employ the same minimum elevation angle, which we assume herein to form the sets of overhead satellites as
\begin{align}
\mathcal{P} &= \braces{\vp\in\bar{\mathcal{P}} : \epsilon(\sfu,\vp) \geq \epsilon_{\mathrm{min}}} \subset \bar{\mathcal{P}}, \\% \qquad
\mathcal{S} &= \braces{\vs\in\bar{\mathcal{S}} : \epsilon(\sfv,\vs) \geq \epsilon_{\mathrm{min}}} \subset \bar{\mathcal{S}}, \label{eq:sec_sat}
\end{align}
where $\epsilon(\sfu,\vp)$ denotes the elevation of a satellite $\vp$ relative to the horizon at a user $\sfu$; we take $\epsilon_{\mathrm{min}} = 35^\circ$ \cite{kuiper} in the results that follow.
Both $\mathcal{P}$ and $\mathcal{S}$ are functions of time, since the satellites overhead a particular user will vary as the constellations progress in their orbit; in \leo systems, satellites are overhead a typical user for about two minutes at most.
We capture this time dimension by simulating the system over 24 hours with a resolution of 30 seconds.

\subsection{Satellites, Ground Users, and Path Loss}

\begin{figure}[t]
	\centering
	\includegraphics[width=\linewidth,height=0.25\textheight,keepaspectratio]{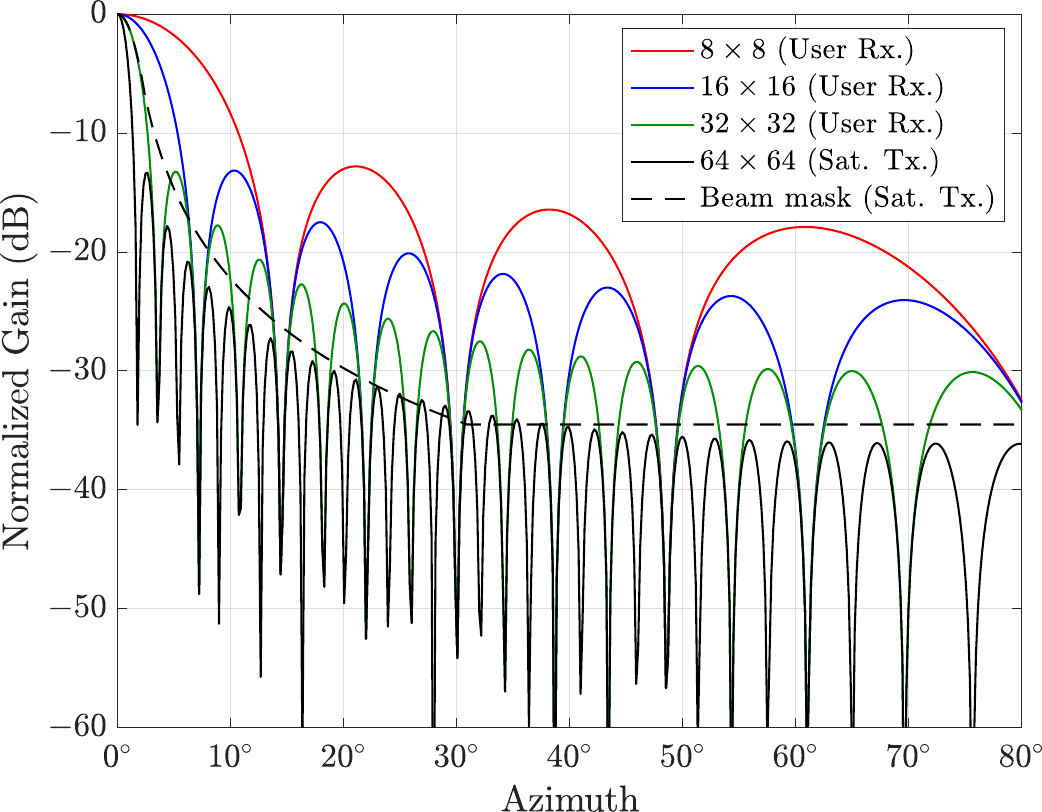}
	\caption{Beam patterns of the 64$\times$64 phased array antenna considered at each satellite and the 8$\times$8, 16$\times$16, and 64$\times$64 phased array antennas considered at each ground user. Shown also is the transmit antenna beam mask for satellites operating at 20 GHz required by ITU \cite{itur1528}.}
	\label{fig:antenna-pattern}
	\vspace{-0.23cm}
\end{figure}	

Having considered the $19.7$--$20.2$ GHz band, we simulate the system at a carrier frequency of $20$~GHz with $400$~MHz of bandwidth.
We assume all satellites and ground users in both the primary and secondary systems are equipped with standard uniform planar antenna arrays with half-wavelength spacing.
We assume satellites are equipped with 64$\times$64 antenna arrays containing 4096 elements and perform canonical matched filter beamforming toward the user they are serving\footnote{For per-cell beamforming, users can be assumed located at the cell centers.}; this delivers a maximum gain that is on par with the $34.5$--$37$ dBi reported in filings \cite{kuiper_ss}. 
The transmit power of each primary satellite is set to a maximum \gls{eirp} of $-54.3$ dBW/Hz and that of each secondary satellite is set to $-53.3$ dBW/Hz, according to \gls{fcc} filings of Starlink \cite{spaceX_ss} and Kuiper \cite{kuiper_ss} when operating in the $20$~GHz band. 
We employ minor power control (factors on the order of $\pm1$ dB) across each constellation to ensure the received signal strength from each satellite is approximately equal; this accounts for minor differences in path loss at slightly different altitudes.

\begin{figure}[t]
    \centering
    \includegraphics[width=\linewidth,height=0.25\textheight,keepaspectratio]{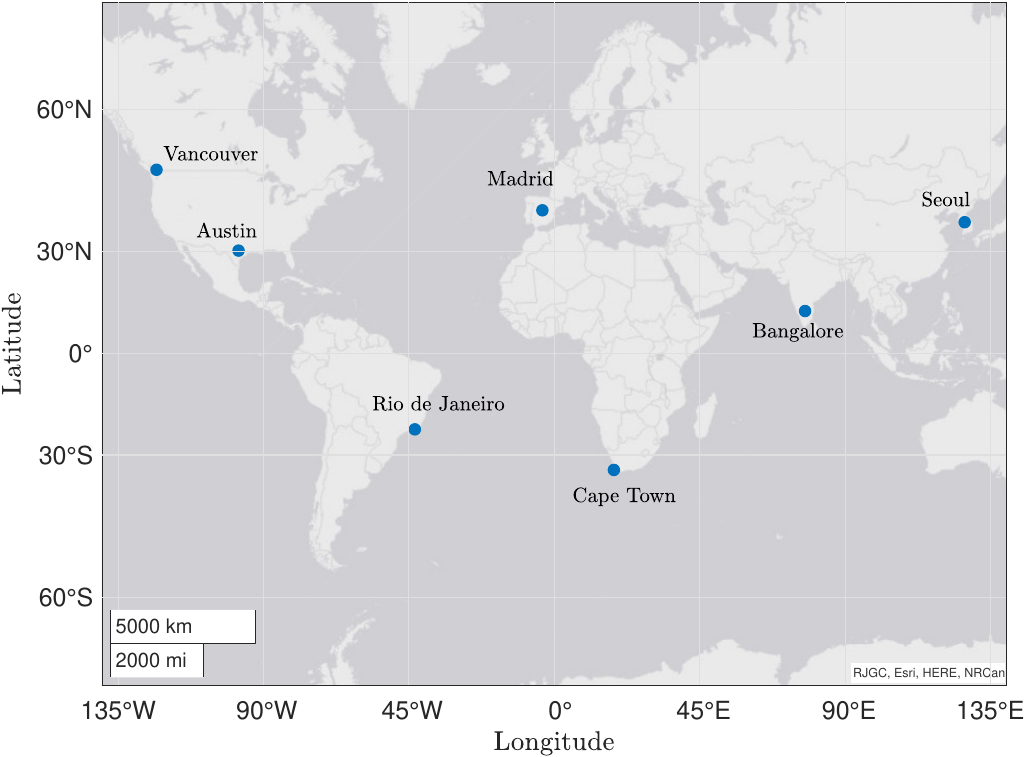}
    \caption{Cities varying widely in latitude from across the globe that are considered in this work. For concreteness, evaluations in Sections~\ref{sec:results-ii}--\ref{sec:results-iii} are specifically for Austin, but similar conclusions are drawn across the globe.}
    \label{fig:cities}
    \vspace{-0.23cm}
 \end{figure}	

We consider ground users equipped with various numbers of antennas, ranging between 8$\times$8, 16$\times$16, and 32$\times$32 antenna arrays. 
In \figref{fig:antenna-pattern}, we show the normalized azimuth cut of the gain delivered by the antenna array at each satellite and each of the ground user's receive antenna arrays under consideration. 
Included in \figref{fig:antenna-pattern} is the transmit beam mask, which is the maximum envelope of the antenna gain that satellites should abide by when transmitting, according to the \gls{itu} \cite{itur1528}; our choice of antenna array at the satellite aligns quite well with this beam mask. 
All ground users are assumed to have a noise power spectral density of $-174$ dBm/Hz plus a $1.2$~dB noise figure, % assuming rooftop-mounted users \cite{3gpp38821}. 
based on recommendations published by 3rd Generation Partnership Project (3GPP) for rooftop-mounted users \cite{3gpp38821}.

In order to assess worst-case interference, we assume the ground users are located at the same point, but empirically we found that even separations of up to 5 km result in the same conclusions drawn, due to the large footprint of a satellite's beam on the surface of the Earth.
\edit{In our simulation, we will examine the interference in multiple cities across the globe, namely Vancouver, Madrid, Seoul, Cape Town, Austin, Rio de Janeiro, and Bangalore, as shown in \figref{fig:cities}. These cities vary widely in latitude, which impacts the number of satellites overhead the users.} 
The path loss $L(\sfu,\vp)$ from a satellite $\vp$ to a ground user $\sfu$ is modeled based on free-space path loss, 
\edit{as atmospheric and scintillation loss in the mid-latitude regions at elevation angles above $35^\circ$ are often observed to be marginal \cite{iturp676, itur_618_13, 3gpp38811} and outdoor user terminals are assumed. 
} 

%\vspace{-0.12cm}
\subsection{Interference Protection Constraint}

The interference inflicted by secondary satellites onto primary ground users is of particular importance from a regulatory perspective. 
The precise definition of \textit{prohibitive interference} is an open question with ongoing discussions and efforts to define such \cite{fnprm}.
In this paper, we investigate one common approach based purely on the strength of interference \cite{ntia_ipc, itur1155, itur1432}. 
As one motivating example, the \gls{itu} defines prohibitive interference as when the effective temperature of the receiver increases by more than $6$\% \cite{ntia_ipc,itur1323} when treating interference as noise. %\cite{1dbsnr}.
In other words, the interference power at a primary user $\sfu$ must not be greater than $6$\% of its noise power $\powernoise(\sfu)$.
Denoting $\powerint(\sfu,\vp;\vs)$ as the interference power inflicted by a secondary satellite $\vs$ onto a primary user $\sfu$ being served by $\vp$, we can directly write this constraint as
\begin{align}
\underbrace{\frac{\powerint(\sfu,\vp;\vs)}{\powernoise(\sfu)}}_{=~\minr(\sfu,\vp;\vs)} \leq 0.06 \ \xrightarrow{\mathsf{in~dB}} \
% \todB{\minr(\sfu,\vp;\vs)} \leq -12.2
\minr(\sfu,\vp;\vs) \leq \underbrace{-12.2~\textrm{dB}}_{\triangleq~\minrth}, \label{eq:ipc}
\end{align}
where we use $\minrth$ to denote the interference threshold, in this case $-12.2~\textrm{dB} \approx 10\ \logten{0.06}$.
We refer to \eqref{eq:ipc} as the \textit{interference protection constraint}, which assume the secondary system is obligated to satisfy.
Because there are active discussions as to what an acceptable level of interference is, this work will examine a few different values for $\minrth$, namely $-15$ dB, $-12.2$ dB, $-6$ dB, and $0$ dB.

\begin{figure}[t]
	\centering
	\includegraphics[width=\linewidth,height=0.25\textheight,keepaspectratio]{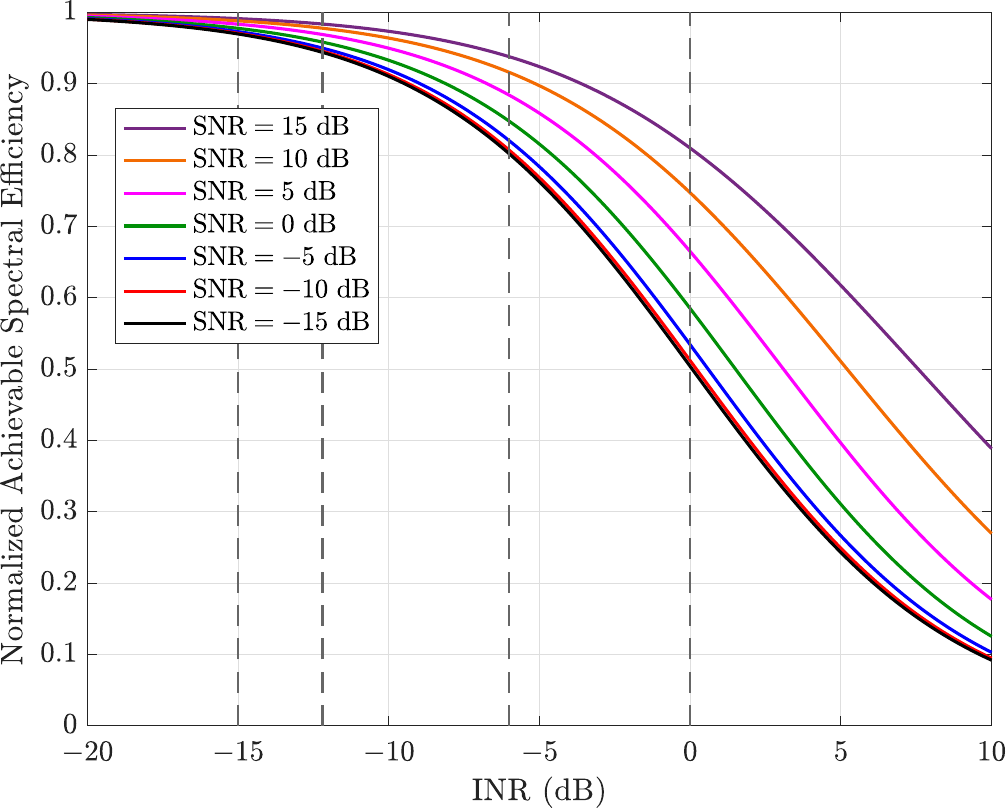}
	\caption{Sacrifices in achievable spectral efficiency \edit{of a single link} due to interference for various \gpsnr. Interference inflicted onto a primary system by the secondary system may only be tolerated if it is below a certain threshold; four thresholds considered in this work are shown as dashed lines. If interference can be kept below $\minr \leq -12.2$ dB, for instance, %a little more than
	\edit{roughly} 5\% is lost in achievable spectral efficiency, at most, even at low \gsnr.}
	\label{fig:inr-vs-capacity}
 % \vspace{-0.25cm}
\end{figure}

Consider \figref{fig:inr-vs-capacity} illustrating the impact of $\minr$ on an arbitrary communication link. %  whose \snr is $\msnr$.
Shown is the degradation in achievable spectral efficiency $\logtwo{1+\msinr}$, normalized to the link's capacity $\logtwo{1+\msnr}$, as a function of $\minr$ for various $\msnr$.
When $\minr \leq -12.2$ dB, the link's achievable rate remains within 5\% of the capacity for \gpsnr as low as $-15$ dB---arguably a justified tax for coexistence.
When interference is as strong as noise at $\minr = 0$ dB, however, \edit{the loss varies greatly with \gsnr and approaches a 50\% reduction at low \gsnr.}
%only under high \gpsnr can the loss be arguably justified.

\section{Bounds on the Interference Inflicted\\onto a Primary Ground User} \label{sec:results-i}

A natural starting point to begin analyzing the feasibility of coexistence is to examine the upper and lower bounds on the amount of interference that secondary satellites may possibly inflict on a primary ground user.

\begin{definition}[Absolute Bounds on INR]
At a given instant, the maximum and minimum interference that can be inflicted onto a ground user $\sfu$ by the secondary system are respectively
\begin{align}
\minrmaxu &=  \max_{\vs \in \setsats} \max_{\vp \in \setsatp} \ \minrups \label{eq:inrmax}, \\
\minrminu &= \min_{\vs \in \setsats} \min_{\vp \in \setsatp} \ \minrups \label{eq:inrmin}.
\end{align}
\end{definition}
These metrics shed light on the severity of interference when coexisting, and looking at their distributions over time will indicate how this severity varies as satellites in both systems progress along their orbits. 
The distribution of $\minrmaxu$, in particular, will \edit{dictate} the need for an explicit interference protection constraint.
For instance, if it is \edit{usually} the case that $\minrmaxu \leq -12.2$ dB, it may be argued that there is no need to consciously protect primary ground users; we will see shortly that this is not the case.

\begin{definition}[Primary Satellite Selection]
While the aforementioned \textit{absolute} bounds are insightful, perhaps more useful are the maximum and minimum interference after the primary system has selected a satellite $\vp\opt$ to serve its user $\sfu$.
Considering it is the burden of the secondary system to protect the primary system, we assume that the primary system performs satellite selection based purely on maximizing its own \gsnr.
Put simply, $\vp\opt$ is defined henceforth as
\begin{align}
\vp\opt = \arg \max_{\vp\in\setsatp} \ \msnrup.
\label{eq:p-sat-selection}
\end{align}
In a real system, satellite selection would certainly involve a multitude of factors beyond \edit{maximizing} \gsnr, but the exact algorithms employed by commercial systems remain proprietary and undisclosed to the public. 
Furthermore, 
our analysis has little dependence on the particular algorithm used in selecting the primary serving satellite. 
\end{definition}

\begin{definition}[Conditional Bounds on INR]
Now, with the assumption that a primary satellite selection $\vp\opt$ has been made to serve $\sfu$, the bounds on interference conditioned on this selection are
\begin{align}
\minrmaxupopt &=  \max_{\vs \in \setsats} \ \minrupopts \label{eq:inrmax-pstar},\\
\minrminupopt &= \min_{\vs \in \setsats} \ \minrupopts \label{eq:inrmin-pstar}.
\end{align}
From the perspectives of both systems, these are presumably more practical bounds since they take into account the important fact, as highlighted in \eqref{eq:inr-p}, that interference at the primary user depends on both the secondary satellite $\vs$ and its primary serving satellite $\vp\opt$.
It naturally follows that $\minrmaxupopt \leq \minrmaxu$ and $\minrminupopt \geq \minrminu$.
\end{definition}

\begin{figure}
	\centering
	\includegraphics[width=\linewidth,height=0.26\textheight,keepaspectratio]{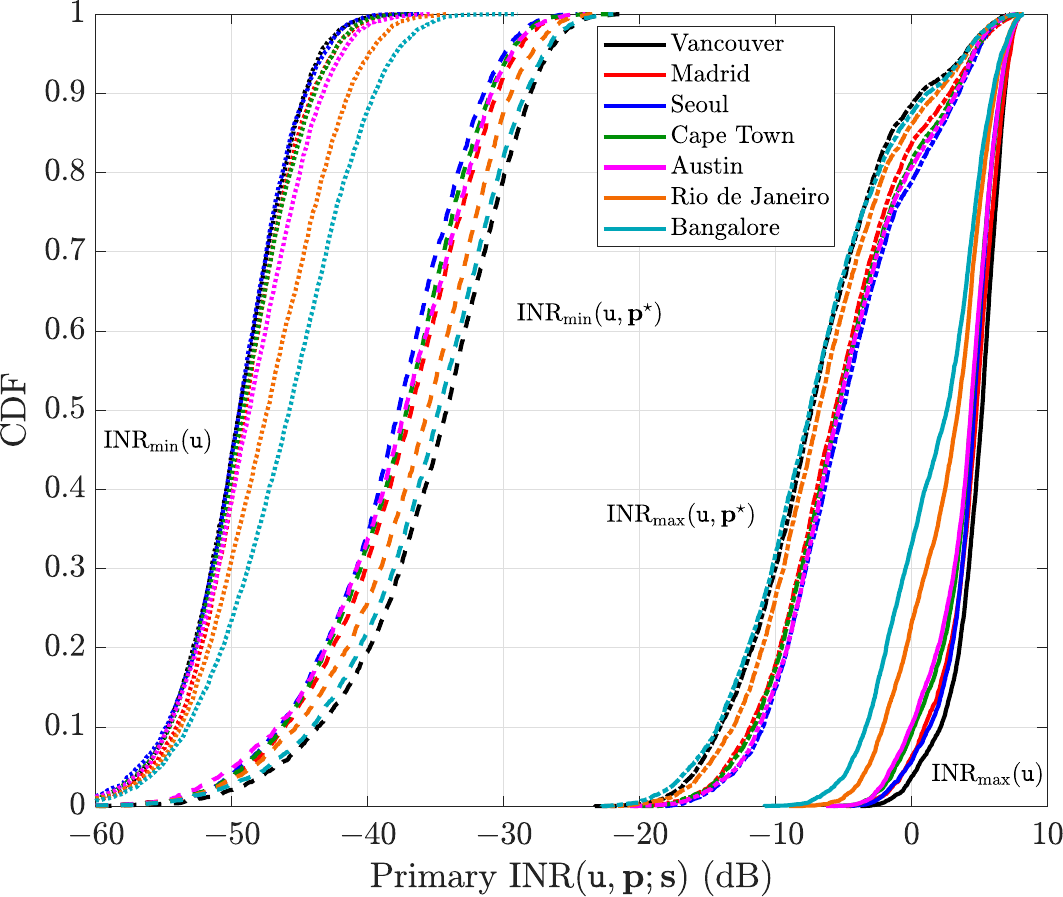} % min_max_pinr_per_antenna.eps}%
	\caption{The empirical CDFs of the absolute and conditional bounds on interference inflicted onto a primary ground user by the secondary system, evaluated at cities across the globe which vary widely in latitude.}
	\label{fig:pinr-globe}
	\vspace{-0.25cm}
\end{figure}

\begin{table}[t]
	\caption{Average Number of Overhead Satellites Across the Globe}
	\centering
	\label{tab:num_sat}
	\begin{tabular}{|c|c|c|c|}
		\hline
		City & Latitude & No.~Primary Sat. & No.~Secondary Sat. \\
		\hline
		Vancouver & $49.2827^\circ$ & 28.29 &10.35  \\
		\hline
		Madrid & $40.4168^\circ$ & 15.37 &16.55\\
		\hline
		Seoul  & $37.5519^\circ$ &13.95  & 18.76 \\
		\hline
		Cape Town &	$-33.9249^\circ$ &12.66 &17.72  \\
		\hline
		Austin & $30.2672^\circ$ & 11.72 &17.39 \\
		\hline
		Rio de Janeiro & $-22.9068^\circ$ & 10.45 &12.98  \\
		\hline
		Bangalore & $12.9716^\circ$ & 9.52 &10.81\\
		\hline
	\end{tabular}
%\vspace{-0.25cm}
\end{table}
%\vspace{-0.12cm}

\edit{
In \figref{fig:pinr-globe}, we plot the empirical \gpcdf of the absolute bounds $\minrmaxu$ and $\minrminu$ and the conditional bounds $\minrmaxupopt$ and $\minrminupopt$ across the globe for primary ground users equipped with 32$\times$32 antenna arrays. %  across cities shown in \figref{fig:cities}.
Each \gcdf is taken over time, and the color of each denotes the users' location.
Let us first examine the absolute bounds.
The maximum interference $\minrmaxu$ is either just below or above the noise floor the vast majority of the time, and we observe that cities at higher latitudes (further from the equator) tend to see higher upper bounds $\minrmaxu$.
This is perhaps best explained by the fact that there simply tend to be more primary-secondary satellite pairs at higher latitudes, illustrated in \tabref{tab:num_sat} and attributed to Starlink's higher concentration of satellites at latitudes around $50^\circ$ \cite{McDowell_2020}.
The steep distributions of $\minrmaxu$ suggest that---at virtually any given time and any location---there is a primary satellite $\vp$ and a secondary satellite $\vs$ in roughly the same direction as one another from the perspective of the primary ground user $\sfu$. 
The minimum interference possible $\minrminu$ is extremely low across all cities, reliably at least $30$ dB below the noise floor. 
In other words, there is virtually always some pair $\parens{\vp\in\setsatp,\vs\in\setsats}$ that can guarantee low interference at a primary user $\sfu$ by virtue of low receive gain $\Grx(\sfu,\vs;\vp) \approx 0$.
Like with $\minrmaxu$, the minimum interference levels $\minrminu$ are also more extreme for higher-latitude users since there are more possible primary-secondary satellite pairs.
}

\begin{takeaway}[There is the potential for extremely high or low interference, depending on the serving satellites] %  depending on the particular primary and secondary serving satellites]
These results on the absolute bounds on interference in \figref{fig:pinr-globe} suggest that there always exists the potential for extremely high interference if certain primary and secondary satellites are serving ground users in the same vicinity.
On the flip side, however, there also always exists the potential for extremely low interference if a certain different pair of satellites is used to serve those same users.
\end{takeaway}

Continuing an examination of \figref{fig:pinr-globe}, we now look at the bounds on interference \textit{after} the primary serving satellite $\vp\opt$ has been selected.
The conditional minimum interference $\minrminupopt$ is very low, more than $20$ dB below the noise floor \edit{across cities} at all times; this suggests that there is always a secondary satellite $\vs\in\setsats$ that can offer low interference, even after the primary serving satellite $\vp\opt$ has been chosen.
Likewise, however, based on the conditional maximum $\minrmaxupopt$, there is always the potential for a secondary satellite to inflict substantial interference that exceeds a threshold of $-12.2$~dB, for example, laid forth before based on \gls{itu} recommendations \cite{itur1323}.

\edit{%
Having conditioned on the primary serving satellite, the variability across cities is largely dictated by the number of secondary satellites overhead, whose average is listed in \tabref{tab:num_sat}.
Cities with more secondary satellites overhead tend to see higher $\minrmaxupopt$ and lower $\minrminupopt$.
In this particular case, Kuiper's constellation has fewer satellites at higher absolute latitudes and near the equator, so these satellites see less extreme $\minrmaxupopt$ and $\minrminupopt$.
In general, the distributions for both $\minrmaxupopt$ and $\minrminupopt$ across cities are fairly similar in shape and both differ in median by a few decibels. 
For the sake of concreteness, our results henceforth will be based on users located in Austin, Texas, with a latitude-longitude of ($30.267153^\circ$, $-97.743057^\circ$), which is where we see %{experiences}
 worst-case $\minrmaxupopt$, along with Seoul.
}

\begin{takeaway}[Strategically selecting a secondary satellite can reliably protect primary ground users]
	Even if the primary system performs satellite selection completely independently of the secondary system, \edit{the distribution of $\minrminupopt$ suggests that} it is still viable for the secondary to strategically select its serving satellite such that low interference is inflicted on the primary ground user. 
	Its ability to do so in a manner that delivers sufficiently high downlink to its own ground user is an open question we examine further in the sections that follow.
\end{takeaway}

\section{System Performance under a Strict Interference Protection Constraint} \label{sec:results-ii}

Beyond solely examining interference, this section also aims to gauge both primary and secondary system performance in terms of \gsinr.
To accomplish this, we will consider various methodologies by which the secondary system selects its serving satellite $\vs \in \setsats$ to serve a ground user $\sfv$.
As with this entire paper, our goal here is not to propose a technique which maximizes system performance but rather to analyze performance under various techniques, with the goal of deriving insights that may drive more sophisticated techniques in future work.

\begin{figure}
    \centering%
    \includegraphics[width=\linewidth,height=0.255\textheight,keepaspectratio]{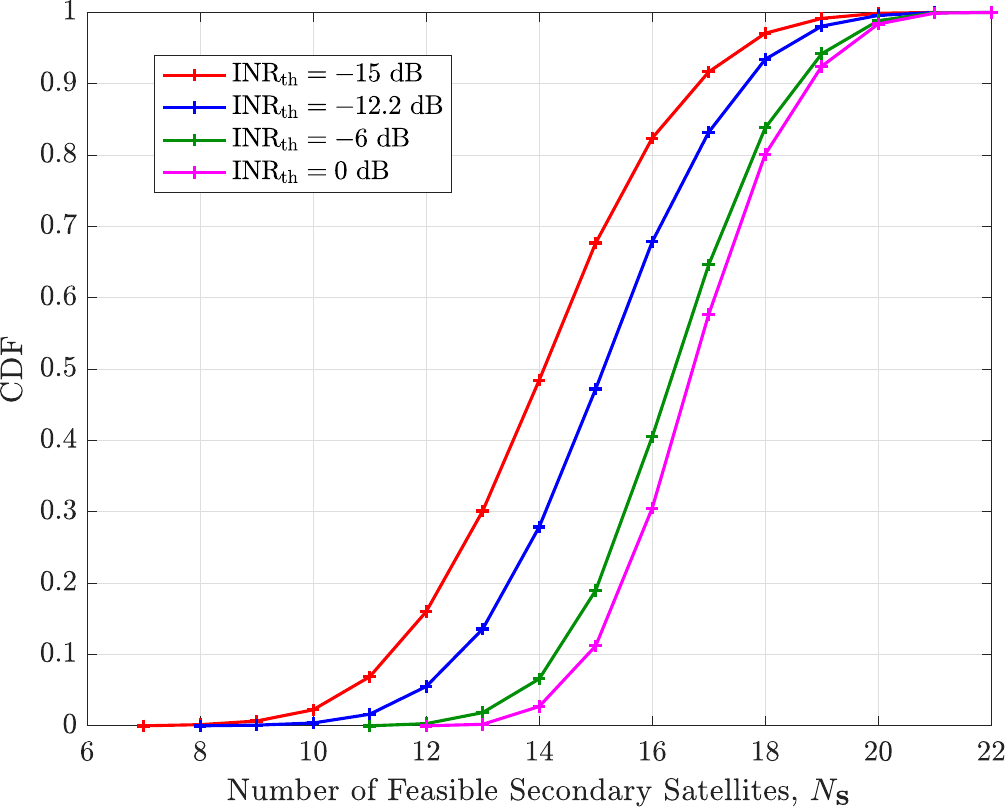}%
    \caption{The empirical CDF (over time) of the number of feasible secondary satellites $N_\vs$ satisfying the interference protection constraint for various thresholds $\minrth$. There are often more than 10 secondary satellites overhead capable of meeting even a very strict protection constraint.}%
    \label{fig:N-feas}%
    \vspace{-0.25cm}
\end{figure}

\begin{figure*}
    \centering
    \subfloat[Primary ground user SINR.	]{\includegraphics[width=\linewidth,height=0.255\textheight,keepaspectratio]{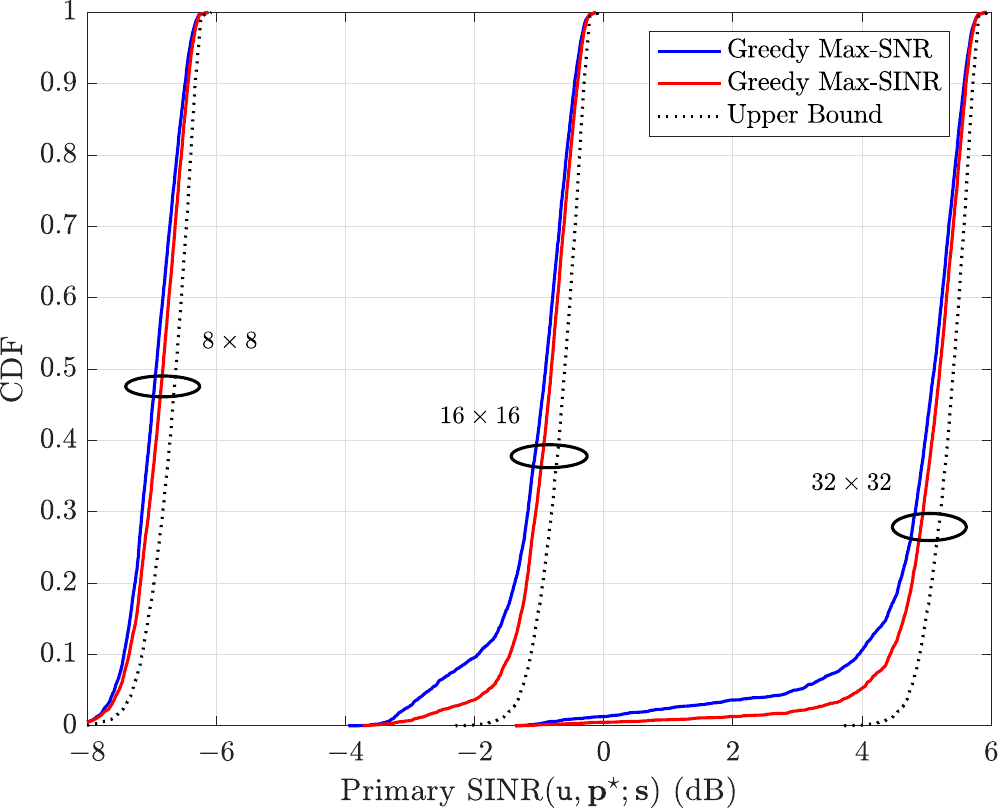}
        \label{fig:cdf-psinr-ant}}  
    \qquad\qquad
    \subfloat[Secondary ground user SINR.
    ]{\includegraphics[width=\linewidth,height=0.255\textheight,keepaspectratio]{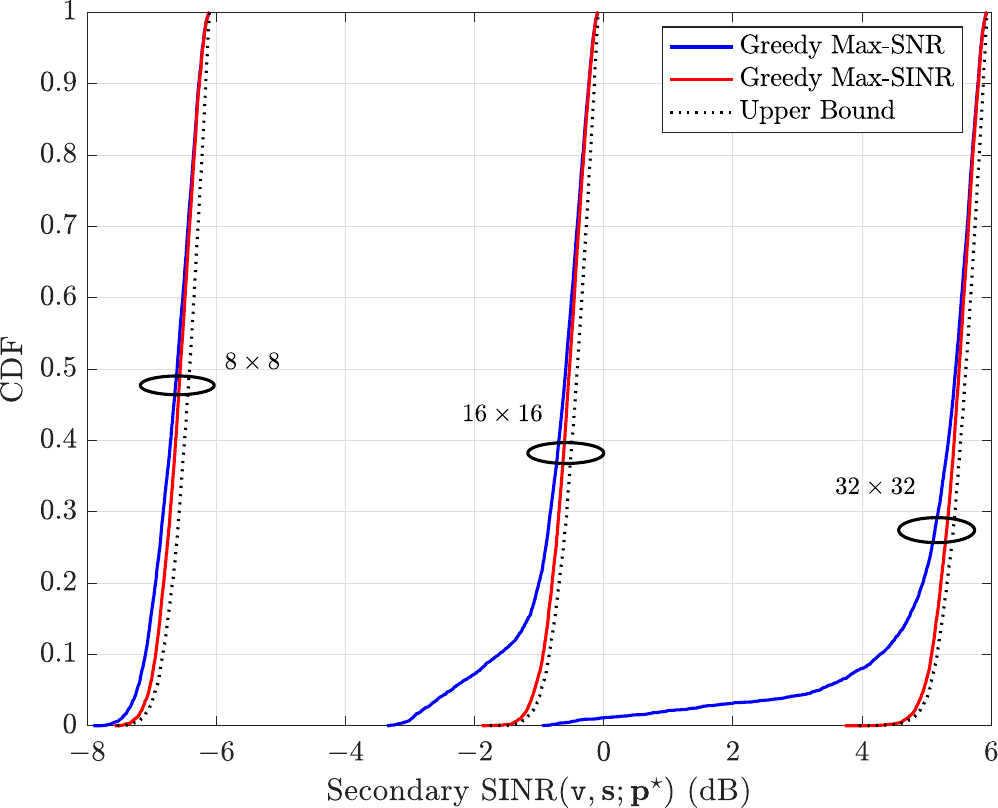}
        \label{fig:cdf-ssinr-ant}}
    \caption{The CDFs of (a) primary SINR and (b) secondary SINR over time under greedy max-SNR and greedy max-SINR secondary satellite selection $\vs_\infty^\dagger$ and $\vs_\infty^\star$ for various ground user antenna array sizes. For each, the upper bound of SINR is shown as a dotted line, corresponding to the interference-free SNR. Choosing a secondary satellite to maximize its SINR inherently also improves the primary SINR. }
    \label{fig:cdf-psinr-ssinr-ant}
    \vspace{-0.25cm}
\end{figure*}

\begin{definition}[Interference Protection Constraint]
When the secondary system selects its serving satellite $\vs \in \mathcal{S}$ in the presence of a primary user $\sfu$ served by a primary satellite $\vp\opt$, imposing an interference protection constraint amounts to this selection satisfying
\begin{align} \label{eq:ipc-2}
\minr(\sfu,\vp\opt;\vs) \leq \minrth.
\end{align}
This is a strict instantaneous constraint in our formulation herein but could take other forms, potentially involving how frequently the threshold is violated, for instance. 
This could make for very interesting \edit{future} work.
\end{definition}

% \subsection{Number of Feasible Secondary Satellites}

\begin{definition}[Number of Feasible Secondary Satellites]\label{def:N-feas}
A natural first question we investigate is the following: How many secondary satellites overhead are capable of satisfying the protection constraint?
This can be formally expressed as the number of feasible secondary satellites $N_\vs$ defined as
\begin{align}
N_\vs = \card{\braces{\vs \in \mathcal{S} : \minr(\sfu,\vp\opt;\vs) \leq \minrth}},
\end{align}
where $\card{\mathcal{A}}$ denotes the cardinality of a set $\mathcal{A}$.
\end{definition}

Recording $N_\vs$ throughout the duration of our simulation as both the primary and secondary satellites orbit allows us to populate its empirical \gls{cdf} in \figref{fig:N-feas} for various thresholds $\minrth$.
It can be seen that over half the time there are at least 15 secondary satellites capable of inflicting interference \edit{less than} %of at most 
$-12.2$ dB when serving their ground user $\sfv$.
In other words, about 55\% of the time, the secondary system has its choice between 15 or more satellites when forced to meet a strict interference constraint of $-12.2$ dB.
This number of feasible satellites reduces with a stricter threshold, whereas relaxing the threshold to $-6$ dB, for instance, adds roughly two satellites to the pool in median.
% Even for a very relaxed threshold of $0$ dB, there are at most 22
In all cases, there are rare but potentially concerning instances where only 6--12 satellites can meet the threshold, imposing limited flexibility in the secondary system's ability to serve its ground user.

% \subsection{Max-SNR and Max-SINR Secondary Satellite Selection}
\begin{definition}[Greedy Max-SNR and Max-SINR Selection]
Let us begin by considering two greedy secondary satellite selection approaches that ignore the interference protection constraint.
We first consider the case where the secondary system purely maximizes its own \gsnr by making its satellite selection as
\begin{align} \label{eq:max-snr}
\vs^\dagger_{\infty} = \arg \max_{\vs \in \setsats} \ \msnr(\sfv,\vs).
\end{align}
Here, the use of $\infty$ can be thought of as representing an infinite interference threshold $\minrth$.
Similarly, we also consider the case where the secondary system maximizes its own \gsinr as
\begin{align} \label{eq:max-sinr}
\vs\opt_{\infty} = \arg \max_{\vs \in \setsats} \ \msinr(\sfv,\vs;\vp\opt).
\end{align}
% where $\star$ is used since \gsinr is a more telling gauge on system performance than \gsnr alone.
It is valuable to consider both of these selections because, given the number of feasible secondary satellites $N_\vs$, it may be the case that the secondary system \textit{inherently} satisfies the interference protection constraint when purely maximizing its own \gsnr or \gsinr.
We will see this is not the case, however.
\end{definition}

In \figref{fig:cdf-psinr-ssinr-ant}, we evaluate the performance of both the primary and secondary systems under the greedy max-SNR and greedy max-SINR secondary satellite selection schemes.
First, let us consider \figref{fig:cdf-psinr-ant}, showing the empirical \gpcdf of primary \gsinr over time under each secondary selection scheme for various receive antenna arrays. 
The upper bounds on \gsinr (i.e., the interference-free \gsnr) for each array size is shown as the dotted line.
For all three array sizes, purely maximizing \gsinr at the secondary system tends to yield primary system performance superior than purely maximizing \gsnr.
Somewhat intuitively, it can thus be concluded that improving the secondary system's \gsinr results in the primary system also enjoying higher \gsinr; in other words, by choosing a secondary satellite that reduces the interference inflicted onto its own ground user $\sfv$ by the selected primary satellite $\vp\opt$, the interference inflicted onto the primary user $\sfu$ is also inherently reduced.
The superiority of max-SINR over max-SNR is most clearly seen in lower tail improvement, which has historically been a point of \edit{concern} for network service providers.

\begin{figure}[t]
	\centering
     \includegraphics[width=\linewidth,height=0.255\textheight,keepaspectratio]{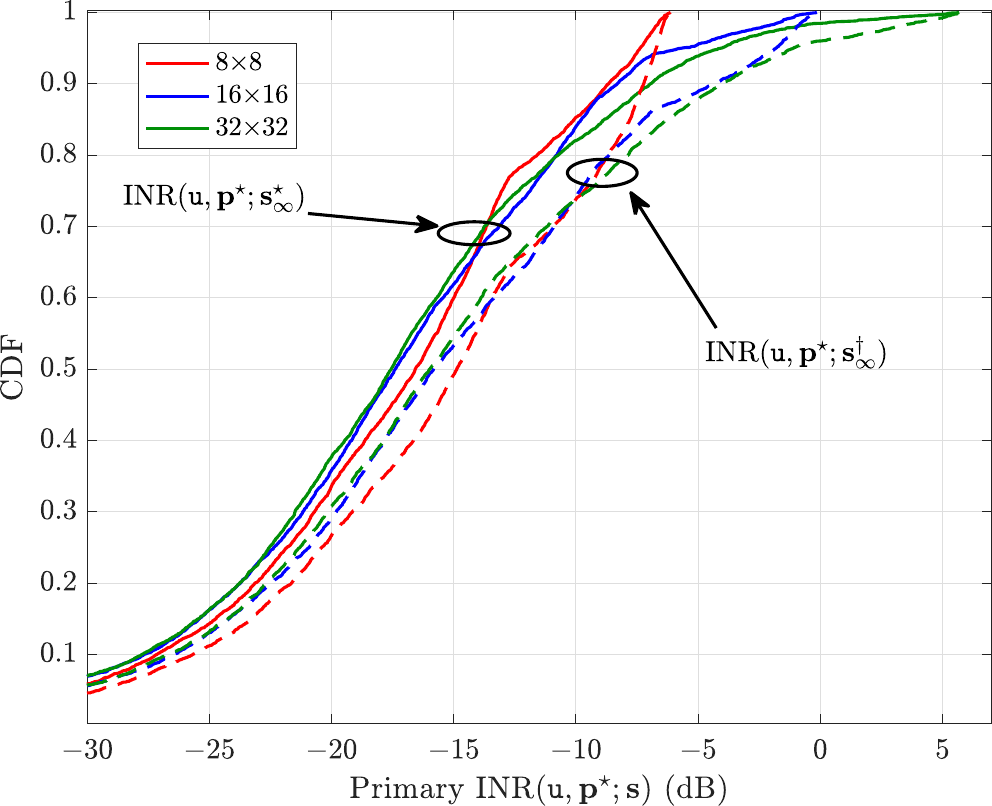}
     \caption{The interference inflicted onto a primary user served by a satellite $\vp\opt$ by a secondary satellite which purely maximizes its own SNR or maximizes its own SINR, without protecting the primary user. Maximizing SINR at the secondary system will inherently reduce interference inflicted onto the primary ground user but not sufficiently so, motivating the need to explicitly incorporate a protection constraint into secondary satellite selection.}
	\label{fig:pinr-obj}  
	   \vspace{-0.25cm}
\end{figure}

Now, we consider \figref{fig:cdf-ssinr-ant}, depicting secondary system \gsinr analogous to \figref{fig:cdf-psinr-ant}.
Many of the same trends can be observed but are more extreme.
The improvement offered by maximizing \gsinr is more obvious as it drastically reduces the density of the lower tail and pulls the distribution quite close to its upper bound.
Clearly, maximizing its own \gsinr would yield the preferred outcome from the perspective of the secondary system but this does not guarantee sufficiently low interference at the primary user at all times.
This can be seen in \figref{fig:pinr-obj}, which compares the primary \ginr under both secondary selection schemes. 
While maximizing the secondary \gsinr results in lower primary interference than maximizing \gsnr, a signification portion of the time, interference is well above the thresholds under consideration (e.g., $-12.2$ dB).
\edit{In \figref{fig:pinr-obj}, we also observe a general increase in the upper tail of interference as the array size increases, since more antennas leads to a higher receive gain in instances when the primary and secondary serving satellites are in the same vicinity.}

\begin{figure*}[t]
    \centering
    \subfloat[Primary ground user SINR.]{\includegraphics[width=\linewidth,height=0.255\textheight,keepaspectratio]{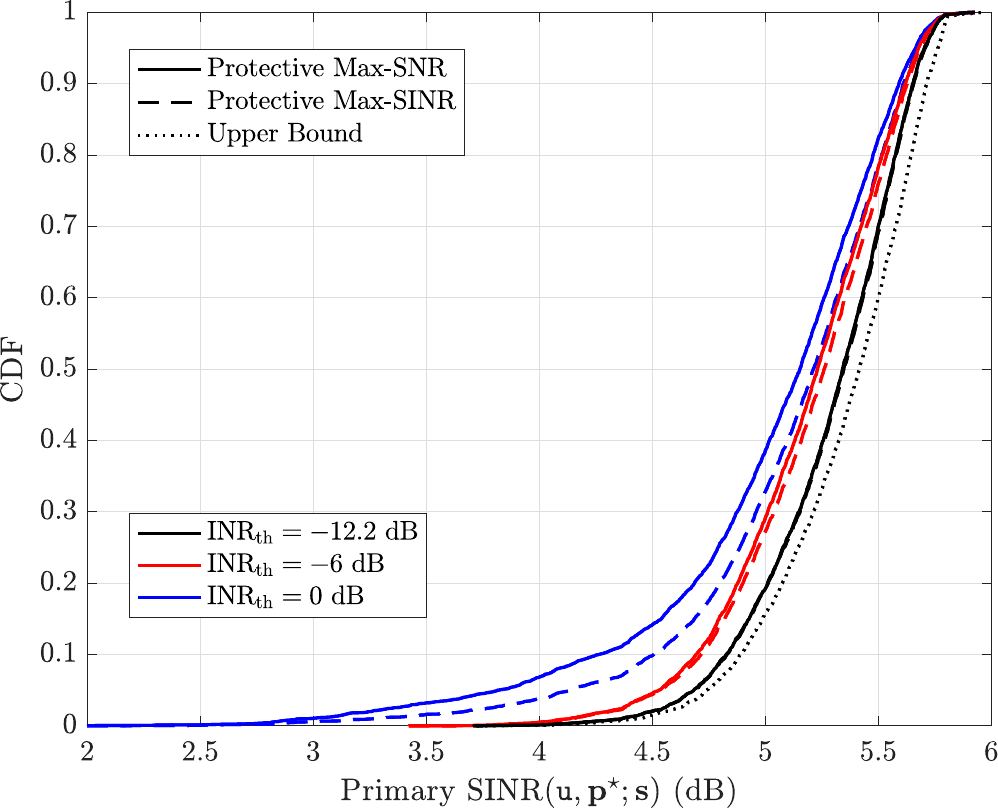}
        \label{fig:cdf-psinr-protect}}
    \qquad\qquad
    \subfloat[Secondary ground user SINR. ]{\includegraphics[width=\linewidth,height=0.255\textheight,keepaspectratio]{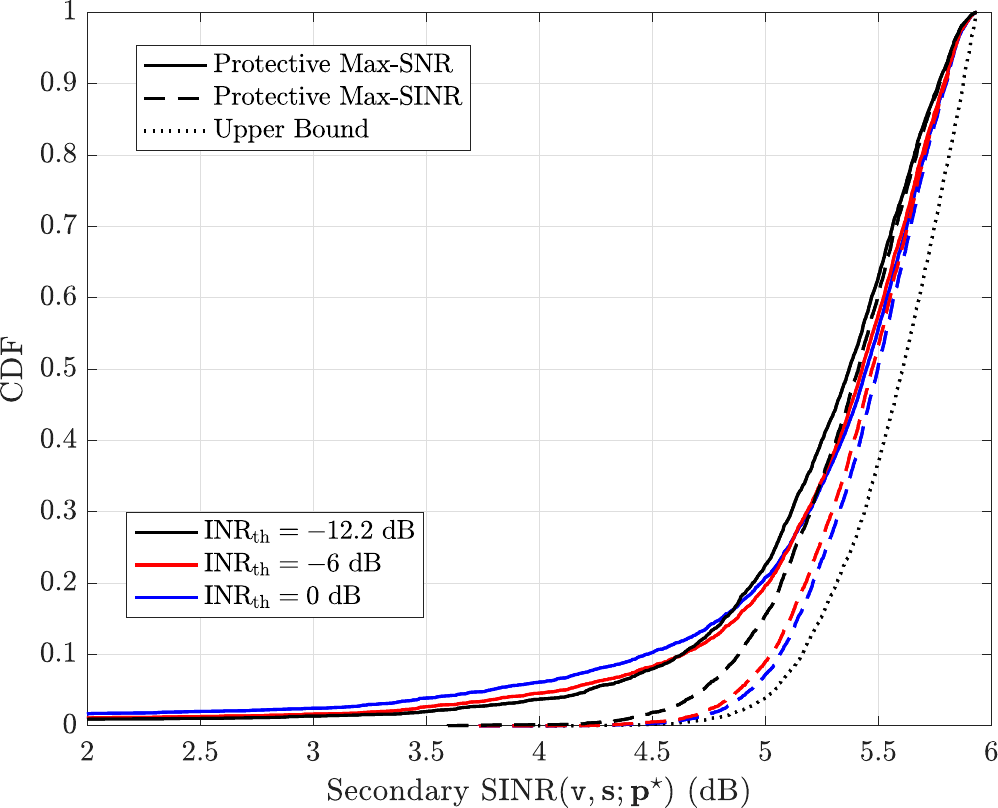}\label{fig:cdf-ssinr-protect}
    }\\	
    \caption{The \gcdf of \gsinr of primary and secondary users with $32\times32$ phased array antennas per \ginr threshold where \ginr threshold is color-coded as black: $\minrth= -12.2$ dB, red: $\minrth=-6$ dB, and blue: $\minrth = 0$ dB. While protecting a primary user, choosing the secondary satellite which maximizes SINR offers far superior secondary and slightly better primary performance than maximizing SNR. 
    }
     \label{fig:cdf-psinr-ssinr-protect}
       \vspace{-0.25cm}
\end{figure*}

\begin{takeaway}[Max-SINR is preferable for both primary and secondary systems, but does not guarantee protection]
The secondary system selecting the satellite which maximizes its SINR is preferred over maximizing its SNR from the perspective of both the secondary system and the primary system, since it will inherently reduce interference at both systems' ground users.
Doing so, however, does not guarantee that the interference inflicted onto the primary system will be below a plausible protection threshold. % sufficiently low for coexistence.
This motivates the need to explicitly protect the primary ground user when selecting the secondary serving satellite.
\end{takeaway}

\begin{definition}[Protective Max-SNR and Max-SINR Selection]
We now augment the previous selection methodologies by enforcing the interference protection constraint \eqref{eq:ipc-2} to form \textit{protective} max-SNR selection
\begin{subequations}\label{eq:max-snr-protect}
\begin{align}
\vs^\dagger = \arg \max_{\vs \in \setsats} \ & \msnr(\sfv,\vs) \\
\st \ & \minr(\sfu,\vp\opt;\vs) \leq \minrth
\end{align}
\end{subequations}
and \textit{protective} max-SINR selection
\begin{subequations}\label{eq:max-sinr-protect}
	\begin{align}
		\vs\opt = \arg \max_{\vs \in \setsats} \ & \msinr(\sfv,\vs;\vp\opt) \\
		\st \ & \minr(\sfu,\vp\opt;\vs) \leq \minrth.
	\end{align}
\end{subequations}
Outright maximizing \gsinr via \eqref{eq:max-sinr-protect} would presumably yield superior secondary system performance over \eqref{eq:max-snr-protect}, but it very well may be the case that meeting the interference protection constraint and maximizing its own \gsnr would inherently improve its own \gsinr.
This makes it interesting to compare these selections and their resulting performance.
\end{definition}

In \figref{fig:cdf-psinr-ssinr-protect}, we compare the performance of both of these protective secondary satellite selection schemes for various interference thresholds $\minrth$ with 32$\times$32 antenna arrays at the ground users.
First and foremost, comparing \figref{fig:cdf-psinr-protect} to \figref{fig:cdf-psinr-ant} illustrates that incorporating the protective constraint improves primary \gsinr, greatly reducing the lower tail and highlighting the importance of explicitly incorporating the protection constraint into secondary satellite selection.
Aside from this, protective max-SINR selection clearly improves primary \gsinr over protective max-SNR, magnifying when $\minrth$ is relaxed.
This is again due to the inherent reductions in interference at the primary user when maximizing the secondary \gsinr; in this case, sometimes those reductions push interference further below $\minrth$, but this is typically only observed when $\minrth$ is less strict.

Now, in \figref{fig:cdf-ssinr-protect}, we derive important conclusions on secondary system performance when abiding by the protection constraint.
The shown upper bound is the interference-free \gsnr. 
The first key observation is that both protective max-SNR and protective max-SINR are capable of delivering appreciable \gpsinr the overwhelming majority of the time.
Protective max-SNR exhibits a heavier tail, since the only interference reduction it enjoys is the inherent reduction when satisfying the protection constraint.
Protective max-SINR, however, impressively sees minor degradation even for the most stringent constraint. 
As the constraint is relaxed, protective max-SINR approaches the upper bound but ultimately falls short due to interference it incurs from the primary system.  

\begin{takeaway}[Protective Max-SINR is preferable for both systems and guarantees primary users are protected]
Protecting a primary ground user does not necessarily impede the secondary system from attaining appreciable \gsinr, falling short of its upper bound by fractions of a decibel.
The gains in secondary system performance by maximizing its \gsinr while meeting the protection constraint are non-negligible compared to maximizing its \gsnr.
As an added benefit, when the secondary system maximizes its \gsinr, the primary system also enjoys an increase in \gsinr.
\end{takeaway}

It is certainly a welcome sight that impressive performance can be achieved while meeting very stringent protection constraints, but it is perhaps more useful to understand how many secondary satellites are capable of such, since systems will naturally be tasked with serving multiple users with multiple satellites. 
We investigate this further in the following.

\begin{figure}
	\centering
	\includegraphics[width=\linewidth,height=0.255\textheight,keepaspectratio]{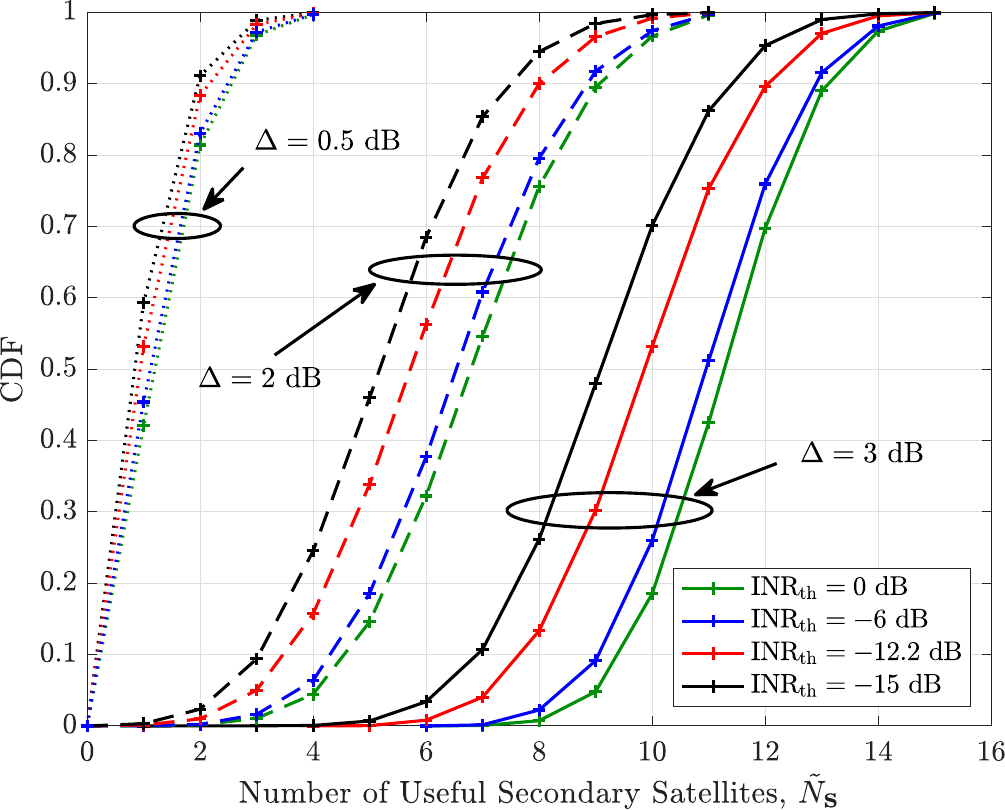}
	\caption{The empirical \gcdf (over time) of the number of useful secondary satellites $\tilde{N}_\vs$ satisfying the protection constraint while offering a secondary \gsinr within a factor of $\Delta$ of the maximum \gsnr. 
	}
	\label{fig:prob-sec-sats}
    \vspace{-0.25cm}
\end{figure}

\begin{definition}[Number of Useful Secondary Satellites]
Definition~\ref{def:N-feas} introduced $N_\vs$, the number of secondary satellites capable of satisfying the protection constraint.
We extend this notion to define the number of \textit{useful} secondary satellites $\tilde{N}_\vs$, capable of satisfying the protection constraint while also offering a secondary \gsinr within a factor of $\Delta$ from the maximum possible \gsnr (i.e., $\msnr(\sfv,\vs^\dagger_\infty)$). %\edit{Or is it the max SINR? } Yes, SNR. 
\begin{multline}
\tilde{N}_{\vs} = \big| \big\{\vs \in \mathcal{S} : \minr(\sfu,\vp\opt;\vs) \leq \minrth, \\ \msinr(\sfv,\vs;\vp\opt) \geq \msnr(\sfv,\vs^\dagger_\infty) \cdot \Delta\inv \big\} \big|
\end{multline}
\end{definition}

\figref{fig:prob-sec-sats} exhibits the empirical \gcdf of the number of useful secondary satellites for various $\Delta$ and various thresholds $\minrth$. 
Earlier in \figref{fig:N-feas}, we saw that there were typically 12--18 feasible secondary satellites, depending on the protection constraint. 
Comparing this to \figref{fig:prob-sec-sats}, we can see that this decreases to 8--12 satellites if we wish to lose at most $\Delta = 3$~dB in secondary \gsinr. 
This decreases further to 4--8 satellites with $\Delta = 2$ dB and to 1--2 satellites with $\Delta = 0.5$~dB.

\begin{takeaway}[Multiple satellites are useful for coexistence at any give time]
While there are often 12--18 secondary satellites which can satisfy even the most stringent protection constraint, there are typically only 1--2 of those satellites which can also deliver near-maximal \gsinr. 
This number grows to 4--12 satellites, however, if $2$--$3$ dB in \gsinr loss can be tolerated.
\end{takeaway}

\begin{definition}[Satellite Angular Separation]
In order to derive real-world meaning from these four different techniques for secondary satellite selection, we introduce the notion of angular separation between two satellites.
Let us define $\angle\parens{\vs_1,\vs_2}$ as the absolute angular separation between two satellites $\vs_1$ and $\vs_2$ with respect to the primary ground user $\sfu$. 
\end{definition}

\begin{figure}[t]
    \centering
    \includegraphics[width=\linewidth,height=0.255\textheight,keepaspectratio]{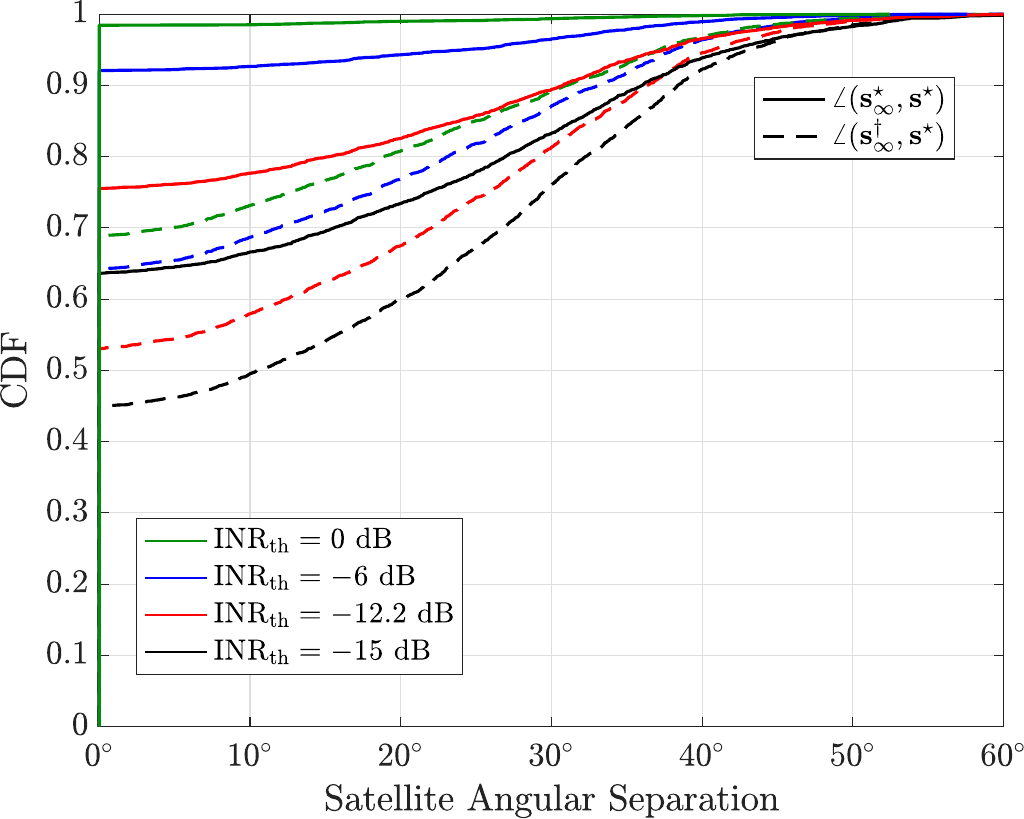}
    \caption{For various $\minrth$, shown are the empirical \gpcdf (over time) of the angular separation between the protective max-SINR secondary satellite selection $\vs\opt$ and (i) the greedy max-SINR selection, $\angle(\vs\opt_{\infty}, \vs\opt)$; and (ii) the greedy max-SNR, $\angle(\vs^\dagger_\infty, \vs\opt)$. The optimal secondary satellite $\vs\opt$ tends to be closer to the greedy max-SINR selection $\vs\opt_\infty$, especially for relaxed $\minrth$.
    }
    \label{fig:angle-dist-max-sinr}
       \vspace{-0.25cm}
\end{figure}

\begin{figure}
    \centering
    \includegraphics[width=\linewidth,height=0.255\textheight,keepaspectratio]{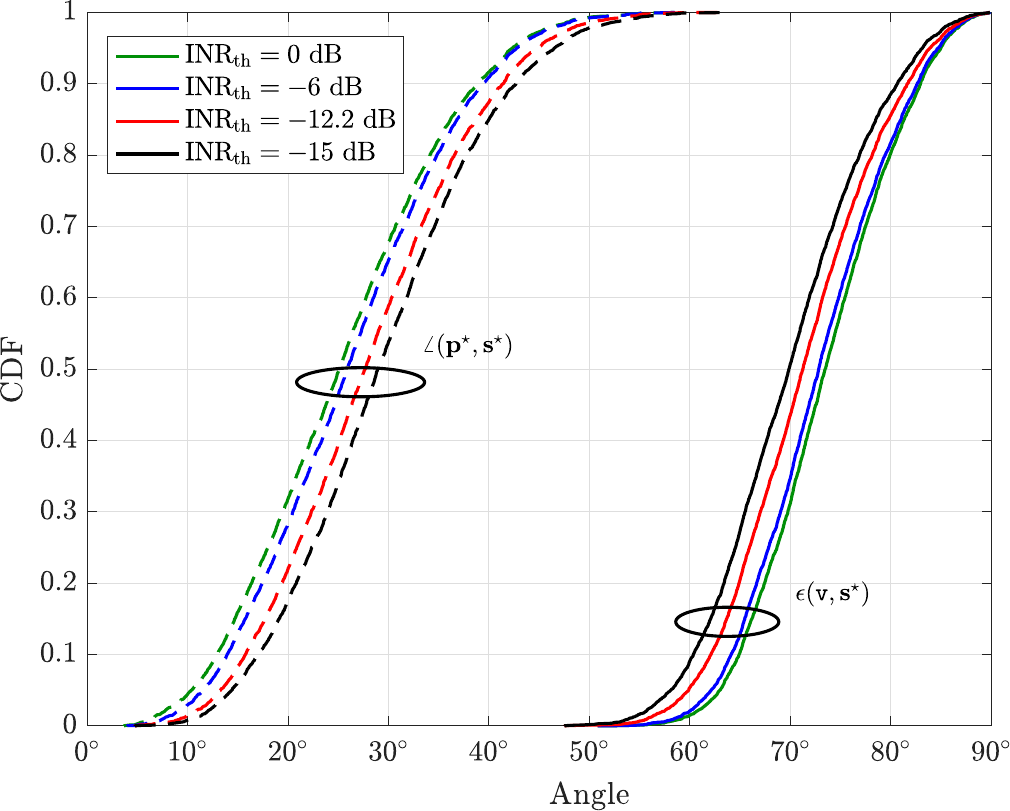}
    \caption{The solid lines depict the empirical \gpcdf (over time) of the elevation angle of the secondary satellite $\vs\opt$ which maximizes its \gsinr while protecting the primary ground user. The dashed lines correspond to that of the angular separation between $\vs\opt$ and the primary serving satellite $\vp\opt$.}
       \label{fig:elevation-angle}
       \vspace{-0.25cm}
\end{figure}

In \figref{fig:angle-dist-max-sinr}, the solid lines depict the distributions (over time) of the angular separation between the protective max-SINR selection $\vs\opt$ and the greedy max-SINR selection $\vs\opt_\infty$ for various $\minrth$.
The dashed lines correspond to the angular separation between the protective max-SINR selection $\vs\opt$ and the greedy max-SNR selection $\vs^\dagger_{\infty}$.
\edit{From the solid green line, we can conclude that about 99\% of the time $\vs\opt = \vs\opt_\infty$; in other words, the protection constraint of $\minrth = 0$~dB is met inherently by simply maximizing the secondary system \gsinr. 
	As the protection constraint is tightened to $\minrth=-12.2$~dB (red solid line), only about 75\% of time does $\vs\opt= \vs\opt_\infty$.
	The remaining roughly $25$\% of the time, the two will typically be separated by $10^\circ$ to $40^\circ$.
	Comparing the solid green line to the dashed green line, we can see that it is far less common yet still frequent that maximizing the secondary system \gsnr also inherently satisfies the protection constraint; as a result, there tends to be more angular separation.
}

Now, in \figref{fig:elevation-angle}, the dotted lines show the angular separation between the primary serving satellite $\vp\opt$ and the protective max-SINR secondary satellite $\vs\opt$.
In general, the two satellites are separated by $15^\circ$ to $45^\circ$, with the separation tending to increase as the protection constraint is made more stringent.
The solid lines depict the elevation angle of the secondary satellite $\vs\opt$. 
With a relaxed protection constraint, the secondary \gsinr is typically maximized by selecting a satellite at higher elevations, typically $65^\circ$ to $85^\circ$.
As the protection constraint is made more stringent, the \gsinr-maximizing satellite is typically at elevations about $5^\circ$ lower, closer to the horizon.

\section{System Performance under Uncertainty} \label{sec:results-iii}

In the previous section, we characterized system performance under secondary satellite selection techniques which protect a primary user $\sfu$ receiving downlink from its serving satellite $\vp\opt$.
Among others, one noteworthy practical challenge in these secondary satellite selection problems is the assumption that the secondary system has knowledge of $\vp\opt$.
This section explores when this is not the case and how such uncertainty may impact system performance.

\begin{figure*}[t!]
    \centering
    \subfloat[Number of feasible satellites under uncertainty.  ]{\includegraphics[width=\linewidth,height=0.26\textheight,keepaspectratio]{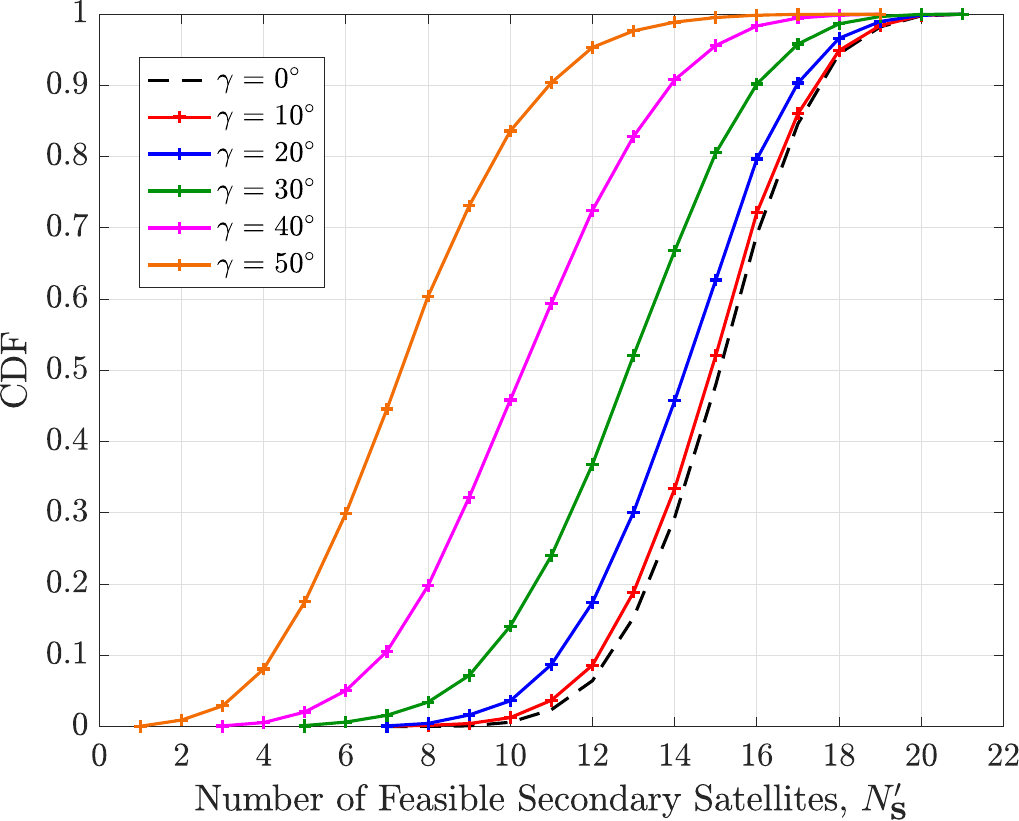}\label{fig:feasible-uncertainty-a}}
    \qquad\qquad
    \subfloat[Average number of feasible secondary satellites under uncertainty. ]{\includegraphics[width=\linewidth,height=0.26\textheight,keepaspectratio]{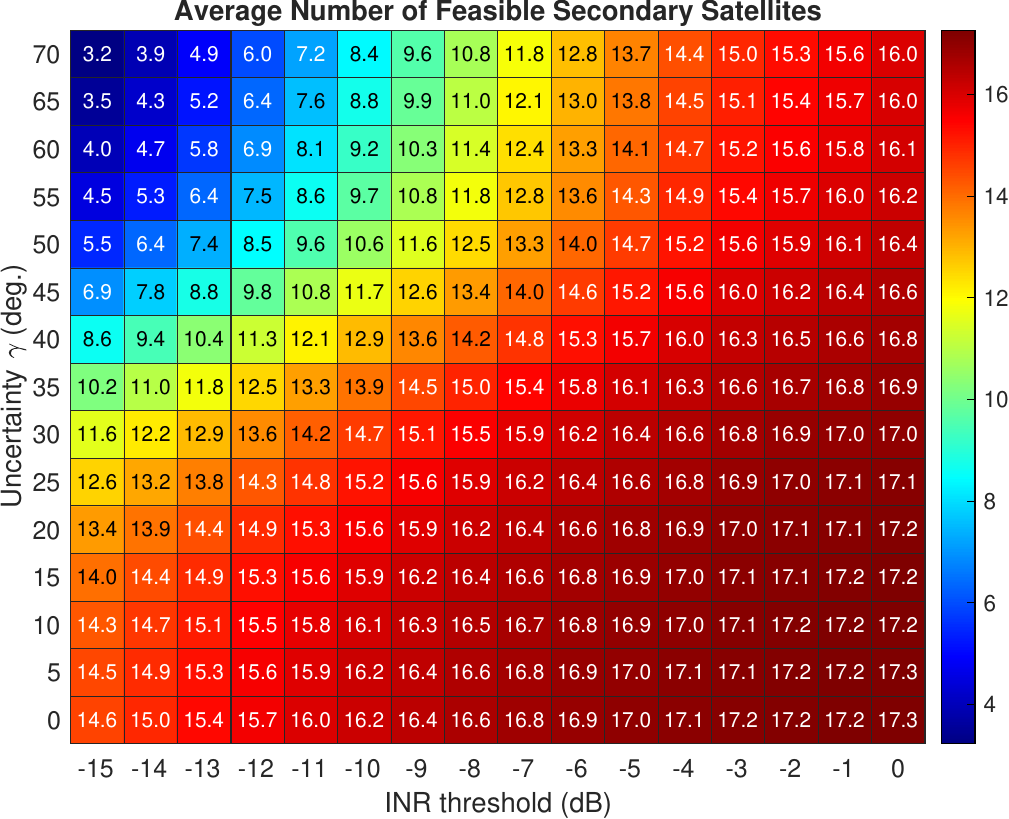}\label{fig:feasible-uncertainty-b}}
    
    \caption{Shown in (a) are the empirical \gpcdf of the number of feasible secondary satellites $N_\vs'$ capable of satisfying a protection constraint of $\minrth = -12.2$~dB for all $\vp\opt \in \setsatp'$ under various levels of uncertainty $\gamma$. In (b), the average number of feasible secondary satellites is shown for various levels of uncertainty $\gamma$ and protection constraints $\minrth$. Even in the most extreme case, there are more than 3 feasible secondary satellites on average.}
    \label{fig:feasible-uncertainty}
   % \vspace{-0.25cm}
\end{figure*}

\begin{definition}[Uncertainty in Primary Satellite Selection]
 Suppose rather than knowing for certain that the primary ground user $\sfu$ is being served by $\vp\opt$, the secondary system instead only knows that the true serving satellite $\vp\opt$ is within some set of primary satellites $\mathcal{P}' \subseteq \mathcal{P}$. 
To model this uncertainty, suppose the secondary system knows (or is confident) that 
\edit{the angular separation of the true primary serving satellite $\vp\opt\in\setsatp$ and some direction $\boldsymbol{\mu}$ is within $\gamma$.
} 
Then, the set of possible serving satellites $\setsatp'$ can be expressed as follows. 
\begin{align}
	\setsatp' = \braces{\vp \in \setsatp : \angle(\boldsymbol{\mu},\vp) \leq \gamma}
\end{align}
As before, the absolute angular difference operation $\angle(\cdot)$ is from the perspective of the primary ground user $\sfu$. 

It is worth emphasizing that, at any given time, it is reasonable to assume that the secondary system has near-real-time knowledge of all primary satellite locations $\setsatp$ since this is public information; this could perhaps be used by the secondary system to populate $\setsatp'$, assuming it can acquire some estimate $\boldsymbol{\mu}$ on the vicinity of the true primary serving satellite.
For the sake of simulation, we construct this set $\setsatp'$ at a given instant by considering all satellites within some angular distance $\gamma$ of the primary satellite which maximizes \gsnr, defined before as $\vp\opt$ in \eqref{eq:p-sat-selection}; in other words, we take $\boldsymbol{\mu} = \vp\opt$.
%\edit{halfway point} between the two users. 
Note that we have not assumed the true primary serving satellite to necessarily be in the direction of $\boldsymbol{\mu}$ nor will it be relevant in the results that follow, since we will focus on \textit{worst-case} secondary system performance under uncertainty. 
\end{definition}

\begin{definition}[Protection Constraint under Uncertainty]
When faced with complete uncertainty about which primary satellite within the set $\setsatp'$ is serving user $\sfu$, ensuring a selected secondary satellite $\vs$ does not exceed a threshold $\minrth$ amounts to the constraint
\begin{align}
% \max_{\vp\in\setsatp'} \ \minr(\sfu,\vp;\vs\opt) \leq \minrth.
\minr(\sfu,\vp;\vs) \leq \minrth \ \forall \ \vp\in\setsatp'.
\end{align}
Naturally, this is a stricter protection constraint than \eqref{eq:ipc-2}, when the secondary system has knowledge of $\vp\opt$, since it must not inflict prohibitively high interference for several potential primary serving satellites.
\end{definition}

\begin{figure*}[t!]
    \centering
    \subfloat[Guaranteed secondary SINR under uncertainty.]{\includegraphics[width=\linewidth,height=0.26\textheight,keepaspectratio]{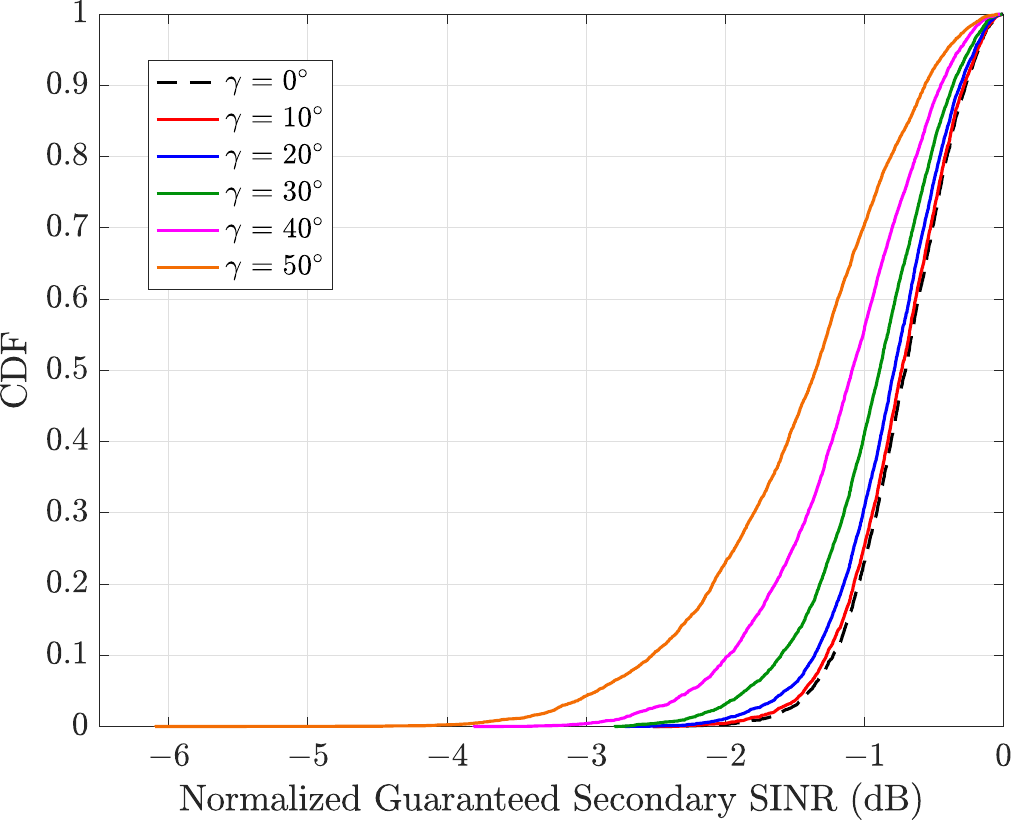}\label{fig:sinr-uncertainty-a}}
    \qquad\qquad
    \subfloat[Median guaranteed secondary SINR under uncertainty.] {\includegraphics[width=\linewidth,height=0.26\textheight,keepaspectratio]{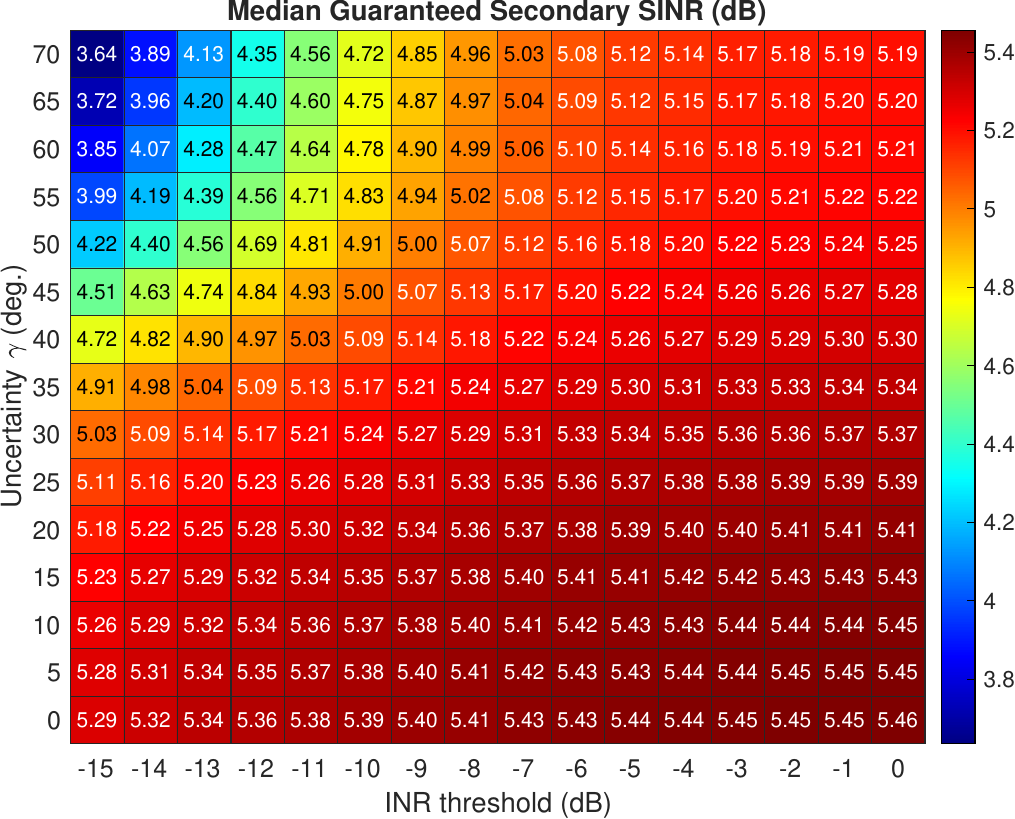}\label{fig:sinr-uncertainty-b}}
    \caption{Shown in (a) are the empirical \gpcdf (over time) of the guaranteed SINR of the secondary system, normalized to its upper bound (\gsnr), under various levels of uncertainty $\gamma$ when $\minrth = -12.2$ dB. In (b), the median guaranteed SINR of the secondary system is shown for various levels of uncertainty and protection constraints $\minrth$. There typically exists secondary satellites which guarantee appreciable SINR while protecting primary users.}
\label{fig:sinr-uncertainty}
\vspace{-0.25cm}
\end{figure*}

\begin{definition}[Number of Feasible Satellites under Uncertainty]
Under uncertainty, the number of secondary satellites satisfying this stricter protection constraint is
\begin{align}
N_{\vs}' = \card{\braces{\vs \in \mathcal{S} : \minr(\sfu,\vp;\vs) \leq \minrth \ \forall \ \vp\in\setsatp' }}.
\end{align}
Naturally, we have $N_{\vs}' \leq N_{\vs}$ as a consequence of uncertainty.
% Evaluating $N_\vs'$ for various levels of uncertainty $\gamma$ will be interesting.
\end{definition}

In \figref{fig:feasible-uncertainty}, we highlight how the level of uncertainty $\gamma$ and the interference threshold $\minrth$ impact the number of feasible secondary satellites $N_\vs'$. 
First, in \figref{fig:feasible-uncertainty-a}, we depict the empirical \gls{cdf} of $N_\vs'$ over time for various $\gamma$ when the threshold is $\minrth = -12.2$ dB.
Notice that the shape of the distribution remains fairly unchanged but undergoes a shift leftward as uncertainty $\gamma$ increases. 
\red{In median, there are typically more than {$4$} fewer feasible satellites when $\gamma = 40^\circ$, compared to when $\vp\opt$ is known exactly, and 8 fewer when $\gamma = 50^\circ$.}
In \figref{fig:feasible-uncertainty-b}, we depict a heatmap of the time-averaged $N_\vs'$ for a wide range of $\gamma$ and $\minrth$.
Even with extreme uncertainty and a very strict threshold, on average there are {$3$} satellites that can guarantee the protection constraint is met across the entire set of possible primary satellites $\setsatp'$.
At $\minrth = -12$~dB, there are more than 10 feasible satellites on average when uncertainty is {$40^\circ$} or less.
With at most $25^\circ$ of uncertainty, the average number of feasible satellites decreases from 17 to 12, compared to a virtually unconstrained and perfectly known scenario (i.e., when $\minrth = 0$ dB and $\gamma = 0^\circ$).

Even though we only assume one secondary satellite is chosen to serve a ground user $\sfv$, it is useful to examine how many satellites are feasible, since it sheds light on the flexibility the secondary system has in meeting the constraint.
For instance, considering a practical secondary system would be tasked with serving multiple users at once across its constellation of satellites, it may benefit greatly from having more satellites capable of meeting a strict protective constraint at any given time.
If $N_\vs' = 1$, for example, the secondary system would be forced to serve the user with the lone feasible satellite or not serve the user at all if the satellite is occupied serving another user.
This also highlights how multi-beam satellites may relax the challenges associated with this process.

\begin{takeaway}[Multiple secondary satellites are feasible under a stringent protection constraint and uncertainty]
	Even with a very stringent interference protection constraint \edit{such as $-12.2$~dB}, there are on average more than 10 secondary satellites capable of protecting a primary ground user under modest knowledge of its serving satellite.
	Likewise, with extreme uncertainty about the primary serving satellite, there are more than 10 secondary satellites capable of protecting a primary ground user under a modest interference constraint.
\end{takeaway}

\begin{definition}[Protective Max-Guaranteed-SINR Selection]
We now investigate the effects of this uncertainty on secondary system performance.
To do so, we introduce the notion of maximum \textit{guaranteed} secondary system \gsinr under uncertainty, which can be accomplished by solving the following constrained satellite selection problem.
\begin{subequations}\label{eq:s-sat-selection-robust}
\begin{align}
% (\hat{\vs}, \hat{\vp})=  \arg \max_{\vs\in\setsats, \vp\in\setsatp'} \  &  \msinrusp \\\ 
(\vs', \vp')=  \arg \max_{\vs\in\setsats} & \ \min_{\vp\in\setsatp'}  \ \msinr(\sfv,\vs;\vp) \\
\st  & \ \minrvps \leq \minr_{\mathrm{th}} \ \forall \ \vp \in \setsatp'
\end{align}
\end{subequations}
Notice that $\msinr(\sfv,\vs';\vp')$ is the minimum SINR that the secondary system will see; we refer to this as the \textit{guaranteed} \gsinr, since for any $\vp \in \setsatp'$
\begin{align}
\msinr(\sfv,\vs';\vp') \leq \msinr(\sfv,\vs';\vp) \ \forall \ \vp \in \setsatp'.
\end{align}
\end{definition}

In \figref{fig:sinr-uncertainty-a}, we show the distribution of guaranteed \gsinr achieved by the secondary system over time, normalized to its upper bound, the maximum \snr (i.e., $\msnr(\sfv,\vs_\infty^\dagger)$ from \eqref{eq:max-snr}), where $\minrth = -12.2$ dB.
\red{By normalizing the guaranteed \gsinr in this way, we can measure the degradation in secondary system signal quality due to the combined effects of (i) abiding by the protection constraint under uncertainty and (ii) interference from the primary system.} 
Without any uncertainty $\gamma = 0^\circ$, there is at worst about a \red{$2.5$~dB} gap in guaranteed \gsinr from its upper bound \red{and less than a $1$~dB gap the majority of the time.}
\red{When $\vp\opt$ is not known precisely, this uncertainty leads to losses in guaranteed \gsinr, especially at the lower tails, but this is most apparent only under high uncertainty.}
\red{For $\gamma \leq 20^\circ$, the guaranteed secondary \gsinr falls short by only a fraction of a dB from that with perfect knowledge of $\vp\opt$.}
\red{When uncertainty increases to $\gamma = 50^\circ$, there is only about a $1$~dB sacrifice in median, but a much more substantial lower tail is present.}
\red{Nonetheless, it is a welcome sight that, the overwhelming majority of the time, less than $4$~dB is sacrificed by the secondary system in protecting (and being interfered by) the primary system, even with very limited knowledge on the primary serving satellite.}

\figref{fig:sinr-uncertainty-b} extends this analysis by depicting the median guaranteed secondary \gsinr \red{(unnormalized)} for various interference thresholds and levels of uncertainty. 
It can be observed in the most extreme case that the minimum guaranteed \gsinr is above $3.5$ dB in median, about $2$ dB short of the upper bound.
In terms of median guaranteed \gsinr, the secondary system is not significantly impacted by making the protection threshold more strict under modest uncertainty. 
When uncertainty exceeds $40^\circ$, however, we begin to see that more stringent $\minrth$ leads to more dramatic losses in median guaranteed \gsinr.

\begin{takeaway}[With limited knowledge about the primary serving satellite, the secondary system can still protect primary ground users and deliver high \gsinr]%[The secondary system can strategically select a satellite, ensuring the protection of the primary user while simultaneously delivering appreciable \gsinr under uncertainty]
Even when the secondary system does not know precisely which primary satellite is serving a particular ground user, if it has limited knowledge on the general vicinity of the primary serving satellite, it can still select a satellite which guarantees the primary user is protected and delivers appreciable \gsinr.
Although, uncertainty leads to wider variability in the \gsinr the secondary system is capable of guaranteeing, \edit{the severity of such is most apparent under fairly high uncertainty}. 
With that being said, the number of feasible satellites does decrease, limiting the freedom a practical secondary system would have in scheduling its satellites when serving multiple ground users and protecting the primary system.
\end{takeaway}

\red{From \figref{fig:sinr-uncertainty-a}, it is certainly remarkable at first glance that the secondary system sacrifices at most only $4$~dB in guaranteed \gsinr under an uncertainty of $\gamma = 50^\circ$ and a strict interference constraint of $\minrth = -12.2$~dB, compared to exact knowledge of $\vp\opt$.} 
The \gsnr delivered by a secondary satellite $\vs \in \setsats$ is dictated mostly by its path loss to the ground user, which itself depends on the satellite altitude and elevation angle \cite{3gpp38821}.
Across all overhead satellites, there is typically \red{$4$--$5$~dB} of variability in \gsnr delivered to a given user \cite{our_tvt_sr}. % , with their altitudes as in \tabref{tab:kuiper} and $\epsilon_{\mathrm{min}}=35^\circ$. 
\red{Combining this with the fact that satisfying the interference protection constraint often inherently reduces interference onto the secondary ground user, a \textit{feasible} secondary satellite typically only sacrifices at most roughly \red{$4$--$5$~dB} in \gsinr when $\minrth$ is strict.}
\red{Selecting the feasible satellite which maximizes \gsinr therefore often results in less than $4$--$5$ dB of loss.}

\vspace{-0.1cm}

\section{Conclusion and Future Directions} \label{sec:conclusion}

This work has investigated the feasibility of in-band coexistence between two heterogeneous \leo satellite communication systems by analyzing two preeminent commercial entities: SpaceX's Starlink as the primary system and Amazon's Project Kuiper as the secondary.
We saw that at virtually anytime, at least one secondary satellite has the potential to inflict prohibitively high interference onto a primary ground user, even with highly directional beams at both the user and satellite.
However, it was also observed that there almost always exists one (or often more) secondary satellites which inflict \edit{acceptable} %substantially lower 
interference onto the primary user while also delivering downlink SINR that approaches its upper bound. 
We showed that this is case even when the secondary system is not certain which primary satellite is serving that particular user.
Based on these results, it can be concluded that in-band coexistence is indeed feasible through strategic satellite selection, but there remain open questions on practical mechanisms to execute such selection.

\edit{Interesting extensions of this evaluation on coexistence would be to cases where each satellite forms multiple beams, where the secondary system has uncertainty about the locations or receivers of primary ground users, and where the secondary system is tasked with scheduling satellites across its entire network.}
Creating novel techniques that leverage satellite selection or other means---potentially harnessing machine learning---to facilitate coexistence between \leo satellite communication systems on a network scale would be extremely valuable contributions.
It would also be useful to formulate and extensively evaluate other interference protection constraints, 
\edit{considering, for example, a probabilistic constraint in time, constraints on handover frequency, and even user mobility.}
Investigating how coexistence may be facilitated by limited cooperation between the primary and secondary systems would make for interesting future work as well.

% \newpage

%\input{sec-bibliography.tex} 

\vspace{-0.15cm}

\section*{Acknowledgments}

We thank Arun Ghosh, Michael Hicks, Anil Rao, and Vikram Chandrasekhar from Amazon's Project Kuiper for their valuable discussions and feedback on our methodology and results.

\vspace{-0.12cm}

\bibliographystyle{bibtex/IEEEtran}
\bibliography{bibtex/IEEEabrv, refs}
%\bibliography{refs}

%\input{sec-appendix.tex}

\end{document}